\documentclass[twocolumn,showpacs,nofootinbib,aps,prd,amsmath,amssymb,nobibnotes,floatfix]{revtex4-1}

\usepackage{hyperref}
\usepackage{amsfonts}
\usepackage{mathrsfs}
\usepackage{epsfig}
\usepackage{graphicx}               
\usepackage{url}
\usepackage{hyperref}
\usepackage{float}
\usepackage{color}

\usepackage{graphicx}
\usepackage{epsf}
\usepackage[utf8]{inputenc}

\usepackage{comment}

\newcommand{\nc}{\newcommand}

\nc{\beq}{\begin{equation}}
\nc{\eeq}{\end{equation}}
\nc{\beqa}{\begin{eqnarray}}
\nc{\eeqa}{\end{eqnarray}}

\usepackage{slashed}

\newcommand{\lsim}{\!\mathrel{\hbox{\rlap{\lower.55ex \hbox{$\sim$}} \kern-.34em \raise.4ex \hbox{$<$}}}}
\newcommand{\gsim}{\!\mathrel{\hbox{\rlap{\lower.55ex \hbox{$\sim$}} \kern-.34em \raise.4ex \hbox{$>$}}}}

\def\be{\begin{equation}}
\def\ee{\end{equation}}

\newcommand\affspc{\vspace{4pt}}
\newcommand{\N}{$N$}

\usepackage{footmisc}
\usepackage{setspace}

\begin{document}

\title{Einstein-Vlasov Calculations of Structure Formation}

\author{William E.\ East$^1$, Rados\l{}aw Wojtak$^{2}$, and Frans Pretorius$^3$}
\affiliation{$^1$Perimeter Institute for Theoretical Physics, Waterloo, Ontario N2L 2Y5, Canada \affspc}
\affiliation{$^2$DARK, Niels Bohr Institute, University of Copenhagen, 
Lyngbyvej 2, 2100 Copenhagen, Denmark\affspc}
\affiliation{$^3$Department of Physics, Princeton University, Princeton, New Jersey 08544, USA \affspc} 

\begin{abstract}
We study the dynamics of small inhomogeneities in an expanding universe
	collapsing to form bound structures using full solutions of the
	Einstein-Vlasov (\N-body) equations. We compare these to standard
	Newtonian \N-body solutions using quantities defined with respect to
	fiducial observers in order to bound relativistic effects.  We focus on
	simplified initial conditions containing a limited range of length
	scales, but vary the inhomogeneities from small magnitude, where the
	Newtonian and general-relativistic calculations agree quite well, to
	large magnitude, where the background metric receives an order one
	correction.  For large inhomogeneities, we find that the collapse of
	overdensities tends to happen faster in Newtonian calculations relative
	to fully general-relativistic ones. Even in this extreme regime, the
	differences in the spacetime evolution outside the regions of large
	gravitational potential and velocity are small.  For standard
	cosmological values, we corroborate the robustness of Newtonian \N-body
	simulations to model large scale perturbations and the related cosmic
	variance in the local expansion rate.
\end{abstract}

\maketitle

\section{Introduction}
Recently, there has been a growing interest in quantifying the importance of
effects that are both nonlinear and relativistic on the large scale evolution
and development of structure in the Universe~\cite{Bentivegna:2015flc,Giblin:2015vwq,Rekier:2014rqa,Giblin:2016mjp,Macpherson:2016ict,Daverio:2016hqi,East:2017qmk,Macpherson:2018btl,Giblin:2018ndw,Daverio:2019gql}. 
This means studying effects that may be missed by
the standard tool for studying cosmological structure formation: Newtonian
\N-body simulations.  The motivation for such studies ranges from answering
claims that small scale nonlinearities may have a strong ``backreaction" on
large scales on the one extreme~\cite{Buchert:1999er,Kolb:2004am,Rasanen:2011ki,Ishibashi:2005sj,Green:2014aga}, 
to the desire to quantify small, subpercent
relativistic effects which may soon become observable in the era of precision
cosmology~\cite{euclid2011,lsst2012,des1yr2017}.

There are a number of challenges in performing a full, nonperturbative general-relativistic (GR) calculation of structure formation. Solving the Einstein
equations requires both solving a set of constraint equations (typically
elliptic) at the initial time and evolving hyperbolic equations for the metric
which have characteristics that propagate at the speed of light. The latter
imposes a severe restriction on the timestep of the simulation compared to the
case where the gravity is completely determined by an elliptic equation and
the matter moves nonrelativistically. Resolving the small scales of collapsed
structures is already very challenging within the Newtonian
framework~\cite{Heitmann:2007hr,Schneider:2015yka}, and this restriction makes
the GR case much more severe.  Hence, most calculations beginning with a range
of length scales very quickly become underresolved.  One approach is to only
include some general-relativistic corrections which do not break the elliptic
description of gravity~\cite{Adamek:2015eda,Barrera-Hinojosa:2019mzo}. However,
this requires making \emph{a priori} assumptions about which terms can be
neglected.

GR simulations also tend to discretize the metric functions on grids,
which makes it natural to use a fluid description of the cold dark matter which
can be discretized on the same grid. This is what has been done for most full
GR calculations of cosmological structure to date
(Refs.~\cite{Pretorius:2018lfb,Giblin:2018ndw,Daverio:2019gql} are exceptions
to this). However, such fluid descriptions break down as soon as multistream
regions emerge, which of course are generic features of structure formation.

Finally, there is the difficulty of distinguishing and quantifying the
magnitude of effects coming from nonlinear gravity, from those solely due to
nonlinear perturbations in the matter (which \emph{will} be captured by
standard Newtonian calculations)~\cite{East:2017qmk}. For example, one cannot simply
look at how inhomogeneous various functions of the metric are in a GR
simulation.  Related to this, when one is considering nonlinear deviations from
a homogeneous spacetime, coordinate ambiguities make it difficult to interpret
the metric functions directly, and one has to be careful to compute gauge
invariant quantities in order to make a meaningful
comparison~\cite{Giblin:2018ndw}.

This work extends that of Ref.~\cite{East:2017qmk}, where a direct comparison
of Newtonian and GR simulations of structure formation was performed utilizing
the dictionary of Refs.~\cite{Chisari:2011iq,Green:2011wc} to generate consistent 
initial conditions in both simulations and to compare observables. In
Ref.~\cite{East:2017qmk}, a fluid description of the matter was used for the GR
calculations, which meant that the comparison became unreliable past the point
where multistream regions would develop. Here, we use the methods of
Ref.~\cite{Pretorius:2018lfb} to solve the Einstein-Vlasov equations, 
allowing us to continue the comparison as bound structures are formed.  We
sidestep some of the computational challenges mentioned above by considering
simplified initial conditions, where the perturbations are concentrated at a
single wavelength, but consider various magnitudes for the inhomogeneities.
For large enough inhomogeneities (in excess of standard cosmological values), we
do find appreciable deviations between the Newtonian and GR calculations, with
the collapse of overdensities happening faster in the former.  However, in the
regime where this occurs, it is already clear from the Newtonian calculation
itself that deviations are expected since the gravitational potential and
velocities relative to the speed of light are becoming comparable to unity.
Furthermore, even in such cases, we find that outside the regions of large
gravitational potential, the agreement between the two methods in observables
like the evolution of the density and the propagation of light is still good.

The remainder of this paper is as follows. In Sec.~\ref{sec:method}, we describe
the initial conditions we consider, the methods we use to evolve in both a full
GR and Newtonian framework, and the diagnostic quantities we use to compare the
two.  In Sec.~\ref{sec:results}, we present the results of our calculations
evolving inhomogeneities of various magnitudes, and in Sec.~\ref{sec:conclusion}
we conclude.  In the appendix, we present results estimating the numerical errors
in our calculations.  We use units with $G=c=1$ throughout.

\section{Methodology}
\label{sec:method}
\subsection{Initial conditions}
Following Refs.~\cite{Bentivegna:2015flc,East:2017qmk}, we consider a simple set of initial conditions
consisting of density perturbations about a homogeneous solution. 
The homogeneous solution is characterized by its initial expansion rate $H_0$, and hence
and density $\rho_0:=8\pi/3H_0^2$, which sets the overall scale.
The perturbations
are taken to be in each of the Cartesian directions with
initial wavelength that is four times the Hubble radius at the beginning of the
calculation. That is, we take the Newtonian density contrast to be
\beq
\delta_N = \sum_i \bar{\delta}_i \sin(k x^i) \ , 
\label{eq:simple_id}
\eeq
with $k=\pi H_0/2$. We introduce a small asymmetry between the different Cartesian directions
by letting $\delta_i=\bar{\delta}(1,\ 0.9,\ 1.1)$, and we consider varying magnitude
density perturbations $\bar{\delta}\times 10^2=0.25$, 0.5, 1, and 5.
The initial velocity is given by the Zel'dovich approximation \citep{Zeldovich1970}
\beq
v^i= H_0 \delta_i \cos(k x^i)/k \ . 
\label{eq:v_zd}
\eeq
These initial conditions have a maximum overdensity at $\left(0,0,0\right)$ and maximum underdensity at 
$\left(\pi/k, \pi/k, \pi/k \right)$.

As described in detail in Ref.~\cite{East:2017qmk}, fully general-relativistic initial data 
are calculated using the dictionary of Refs.~\cite{Chisari:2011iq,Green:2011wc} to determine the approximate metric
and stress-energy tensor, and then solving the full Einstein constraint equations
in the conformal thin-sandwich formulation~\cite{idsolve_paper} for any nonlinear corrections.  

\subsection{Newtonian simulations}
The Newtonian \N-body simulations are performed using the \emph{GADGET-2} code
\cite{Springel2005} with a TreePM algorithm for the gravity solver
\citep{Xu1995}.  These simulations serve as a reference to standard
computational cosmology, where the evolution of the cosmic density field is
governed by Newtonian gravity, and is fully separated from the background
expansion, described in turn by the Friedmann equation. GADGET-2 has been
validated in a number of comparison studies verifying the accuracy and
robustness of various numerical implementations of cold dark matter
cosmological simulations (see, e.g., Refs.~\cite{Heit2008,Kim2014,Hei2005}). 

We generate conditions by displacing particles from a regular grid according to the 
field given by the Zel'dovich approximation \citep{Zeldovich1970}
\begin{equation}
\delta x^{i}=-\frac{4\pi}{\rho_{0}}\partial_{i}\Psi_{N}(a=1) \ ,
\end{equation}
where $\Psi_{N}$ is the Newtonian gravitational potential given by
\begin{equation}
\partial^{i}\partial_{i}\Psi_{N}=4\pi a^{2}\rho_{0}\delta_{N} \ ,
\end{equation}
and by convention the scale factor $a$ is set to unity at the beginning
of the calculation.
The resulting density field that is inferred from the positions of the
particles reproduces the input density up to the second order corrections in
the density contrast. As in Ref.~\citep{East:2017qmk}, we apply the corrections
by means of a minimal adjustment of particle's masses. The particle masses are
set in such a way that they compensate all local differences between the actual
(as calculated by the employed density estimator, described below) and input
density evaluated at the position of every particle. We note that the
introduced corrections are small (subpercent level), but they guarantee a
high-accuracy match between initial conditions of the Newtonian and GR simulations.

The density field is not explicitly evolved in the \N-body simulations, and it
can only be derived from the positions of the particles.  Here, we employ a
well-tested method for measuring matter density in cosmological simulations of
cold dark matter, based on tracing the evolution of the Lagrangian tessellation
of the dark matter manifold in phase space
\citep{Shandarin2012,Abel2012}.  Density is estimated by means of scaling the
initial density according to a relative change of the volume of tetrahedral
mass elements defined in the initial tessellation. In single-stream regions (no
shell crossing), local density at a given position is determined solely by a
single tetrahedral cell containing this point, while density in multistream
regions (after shell crossing) arises from multiple density contributions coming
from all locally overlapping tetrahedral cells.

The employed density estimator outperforms more traditional techniques such as
cloud-in-cell (CIC) in several respects. Here, we emphasize that the estimator
can be applied locally, and it does not suffer from undersampling in
single-stream regions, making it an ideal method for tracing the density field
in voids. On the other hand, density estimates in multiple-stream regions should be
regarded with reservation, because the full robustness of the estimator
requires simulations with a computationally heavy adaptive refinement of
tessellation cells \citep{Hahn2016}. In particular, density estimation in the center 
of dark mater haloes depends on resolution, and there is no guarantee that the 
computation can converge due to the cuspy nature of dark matter density profiles, 
although precision estimated from comparing results based on different tessellations 
at fixed resolution is of the order of 0.1 dex \citep{Abel2012}. The problem of resolution 
dependence can be circumvented by employing a density estimator with a fixed smoothing 
scale in comoving coordinates instead. Bearing this in mind, we include CIC estimates of density 
in some cases for comparison with the GR calculation (which does not utilize tetrahedral cells).

Unless otherwise stated, the results shown here are obtained used $N=196^3$
particles. We also run select cases using $N=128^3$ in order to estimate
numerical errors. The simulations were carried out with a force softening 
of $5\times10^{-4}$ (high resolution) and $8\times10^{-4}$ (low resolution) 
in units of the simulation domain length $L$.

In order to compute the trajectories of freely falling test particles, we
follow the evolution of the tetrahedral cells containing the initial positions
of the test particles. The positions of the evolved test particles are then
computed by interpolating between the displacements of cells vertices, which
are always given by dark matter particles.

\subsection{GR simulations}
The fully general-relativistic \N-body simulations are performed using the
methods described in Ref.~\cite{Pretorius:2018lfb}. This code was also recently
used to follow black hole formation from collisionless
matter~\cite{East:2019bwu}.  As in the Newtonian simulations, we determine the
initial particle positions by starting from a uniform lattice of particles and
then displacing each particle slightly according to the Zel'dovich
approximation (given by Eq.~(31) in Ref.~\cite{Chisari:2011iq}).  However,
there will be a small nonlinear correction to the density field which we will
need to apply to the particle distribution. To do this, we use slightly
nonuniform masses for the particles, given by rescaling the masses in
proportion to the ratio of the desired density to that obtained from the
Zel'dovich approximation.

Though the code used here does implement adaptive mesh refinement
(see Ref.~\cite{Pretorius:2018lfb}), for this study we restrict to uniform grids.
We do this mainly for efficiency, though we note that the results in the
appendix indicate that, at late times in our simulations, the numerical error
is mainly dominated by the number of particles.  For most of the results
presented here, we use resolution with 96 points across the wavelength of the
initial perturbation, and $4^3$ particles per grid cell.  However, we run
select cases at multiple resolutions utilizing $2/3\times$ and $4/3\times$ as
many grid points in order to establish convergence and estimate truncation
error. See the appendix for details.

For comparison, we also include a few results that are calculated by treating the matter
as a pressureless fluid as described in Ref.~\cite{East:2017qmk}.

\subsection{Comparing observables}
In order to compare the results of the Newtonian and GR \N-body evolutions, we
compute several quantities defined with respect to fiducial observers, as
detailed in ~\cite{East:2017qmk}.  We compute the matter density along the
worldlines of timelike observers and use this quantity as a function of proper
time $\rho(\tau)$ to define an effective density contrast:
\beq
\delta_{\rm obs}(\tau):=(\rho(\tau)/\rho_0)a_p^{-3}-1,
\eeq
where 
\beq
a_p:=\left[3\tau H_0/2+1\right]^{2/3}
\eeq
is a convenient parametrization of the proper time using the Lema{\^i}tre-Friedmann-Robertson-Walker (LFRW) expression for the scale factor
that would hold in the homogeneous case.
We emphasize that since $H_0$ (and hence $\rho_0:=8\pi/3H_0^2$) is just a constant that
sets the overall scale of our initial conditions, $\delta_{\rm obs}(a_p)$ is just a convenient 
reparameterization of density as a function of proper time.

We also measure properties of the spacetimes using null geodesics which are ``emitted"
and subsequently ``observed" by fiducial timelike observers. If $k^a$ is the four momentum
of the null geodesic and $u^a$ is the four velocity of emitter/observer, we can compute a 
redshift factor
\beq
z = -1+\frac{(u_a k^a)_{\rm emit}}{(u_a k^a)_{\rm obs}}.
\label{eqn:z}
\eeq
For $u^a$, we take the four velocity implied by the stress-energy tensor $T^{ab}=\rho u^a u^b$, which
weights the contributions from different particles in the case of multistream regions.
We can also use the deviation of neighboring null geodesics to compute the luminosity distance
(or, equivalently through the reciprocity relation, the angular distance~\cite{Etherington2007})
as a function of the redshift $D_L(z)$ along each null ray.

For the GR simulations, these quantities are computed by including extra tracer
particles which are evolved in the same way as the matter particles (but
without backreacting). For the Newtonian simulations, these quantities are
computed by reconstructing the effective spacetime using the Newtonian-GR
dictionary of~\cite{Chisari:2011iq,Green:2011wc} and integrating the resulting
geodesic equation.  Hence, the Newtonian calculation also includes relativistic
effects in the propagation of light, etc., and the comparison is really of how
much the spacetimes implied by the two methods of calculation differ.

\section{Results}
\label{sec:results}
\begin{figure}
\begin{center}
\includegraphics[width=\columnwidth,draft=false]{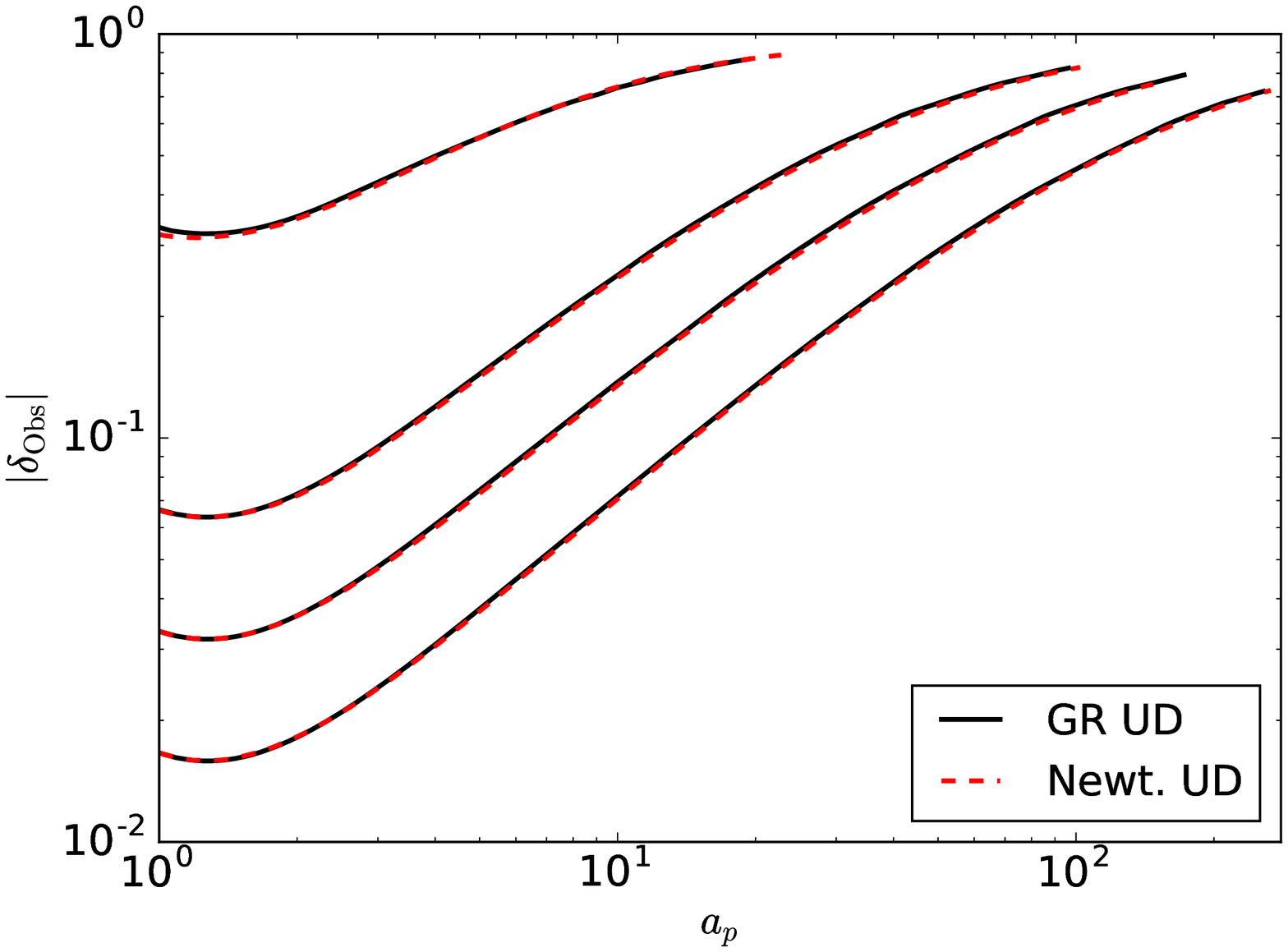}
\includegraphics[width=\columnwidth,draft=false]{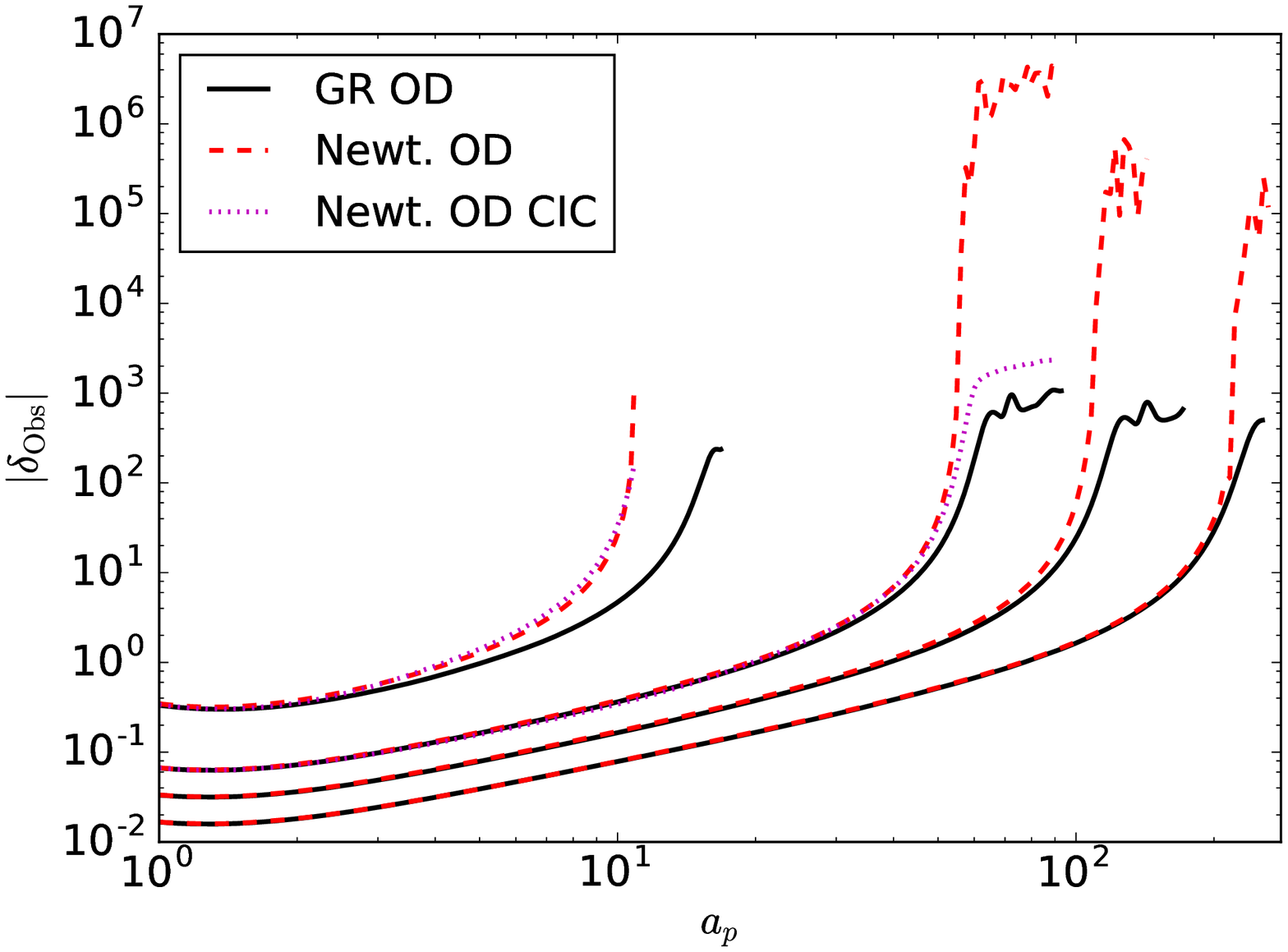}
\includegraphics[width=\columnwidth,draft=false]{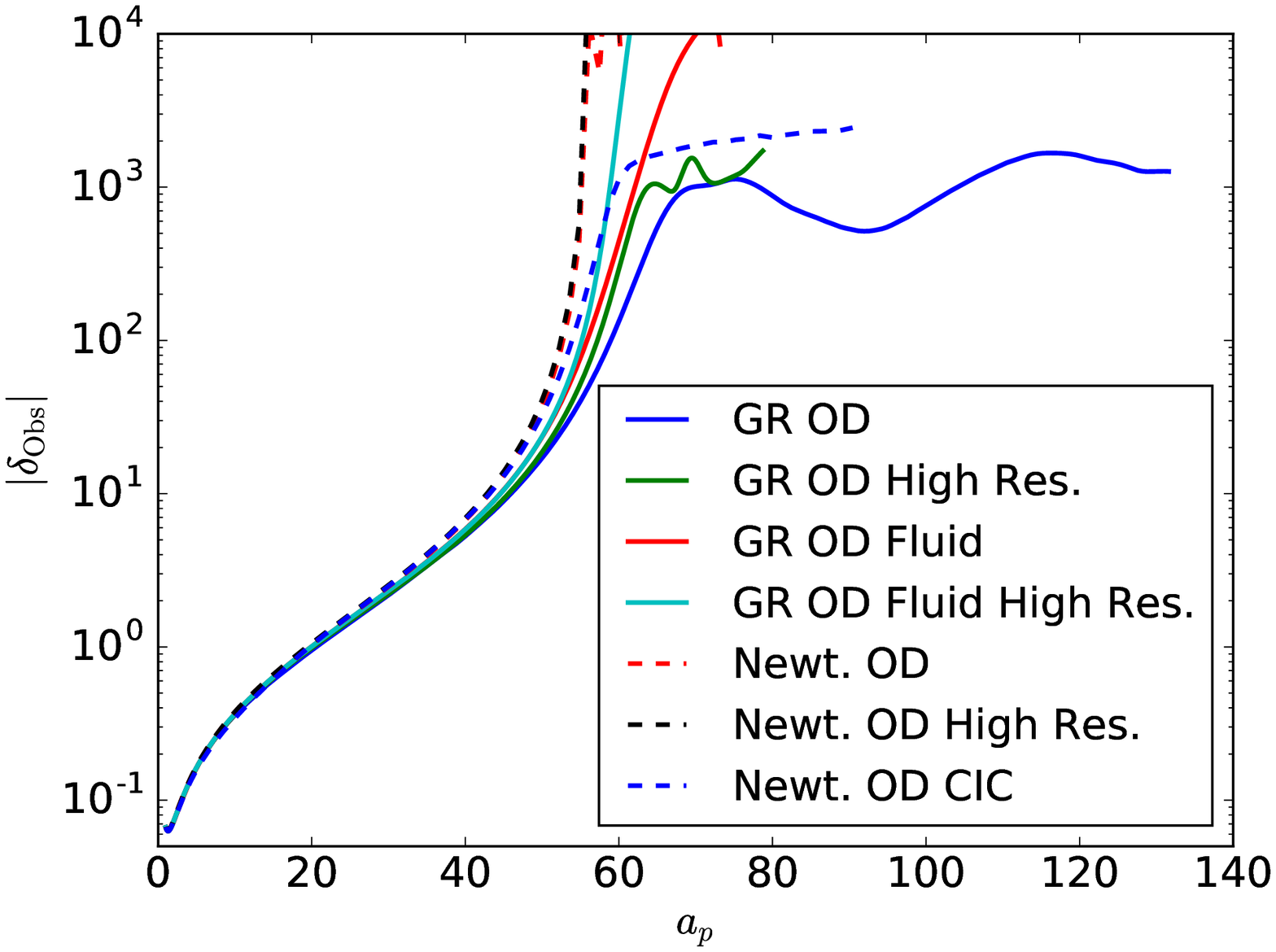}
\end{center}
\caption{
    Top: the $\delta_{\rm obs}$ measure of the density contrast at the points of 
    minimum density for the cases $\bar{\delta}=0.0025$,
    0.005, 0.01, and 0.05.
    (The Newtonian and GR curves for the underdensities are essentially indistinguishable on
    the scale of the plot.)
    Middle: same as above, but for the density contrast at the points of maximum density.
    The curves labeled ``CIC" use a cloud-in-cell estimate of the density---similar
    to the way the calculation is done for GR simulations---instead 
    of the tetrahedral cell estimate. 
    Bottom: a comparison of this quantity at the point of maximum overdensity for $\bar{\delta}=0.01$
            for several different resolutions and utilizing a fluid versus particle treatment.
\label{fig:od_ud}
}
\end{figure}

With the initial conditions we have chosen, as the spacetime expands and the
inhomogeneities move inside the horizon, a growing void emerges at the point of
maximum underdensity, and a bound, multistream region (i.e. a halo) is formed
at the point of maximum overdensity.  In the top and middle panels of Fig.~\ref{fig:od_ud},
we show the density contrast measure $\delta_{\rm obs}$ at these two points for
cases with different magnitudes of the initial inhomogeneities.  The Newtonian and
GR calculations show good agreement at the underdensity for all cases, even as the
density contrast becomes highly nonlinear. 

For the overdensity, two differences are noticeable. The first is that the
collapse and halo formation occurs slightly earlier for the Newtonian case, and
this difference increases as the initial inhomogeneities become larger (and hence
more relativistic). The second is that the saturation density is significantly
larger for the Newtonian case. We shall not focus too much on the latter since
this is fairly sensitive to numerical effects such as the finite number of
particles and the smoothing length. 
In the bottom panel of
Fig.~\ref{fig:od_ud}, we show for the $\bar{\delta}=0.01$ case a comparison of
how this quantity changes, both with numerical resolution, and with a particle
versus pressureless fluid treatment of the matter.
Here it can also be seen that with a CIC estimate of the density, the maximum
density contrast for the Newtonian calculation is much closer to the GR result
(which similarly deposits each particle's stress energy on neighboring grid points).
In the GR pressureless fluid treatment, the calculation breaks down at shell crossing, whereas
with the particle treatment the density eventually saturates. In either case, 
finite resolution tends to lead an underestimate of the density around this point.
However, even taking this into account, the collapse happens faster in the Newtonian
case compared to the GR case.
This discrepancy increases with increasing inhomogeneity amplitude and becomes
quite pronounced for the case with $\bar{\delta}=0.05$.
For this extreme case, the Newtonian
calculation has to be terminated when the magnitude of the Newtonian potential $\psi$ becomes
$\sim 1/2$. We discuss this case in more detail below.

\begin{figure*}
\begin{center}
\includegraphics[width=\columnwidth,draft=false]{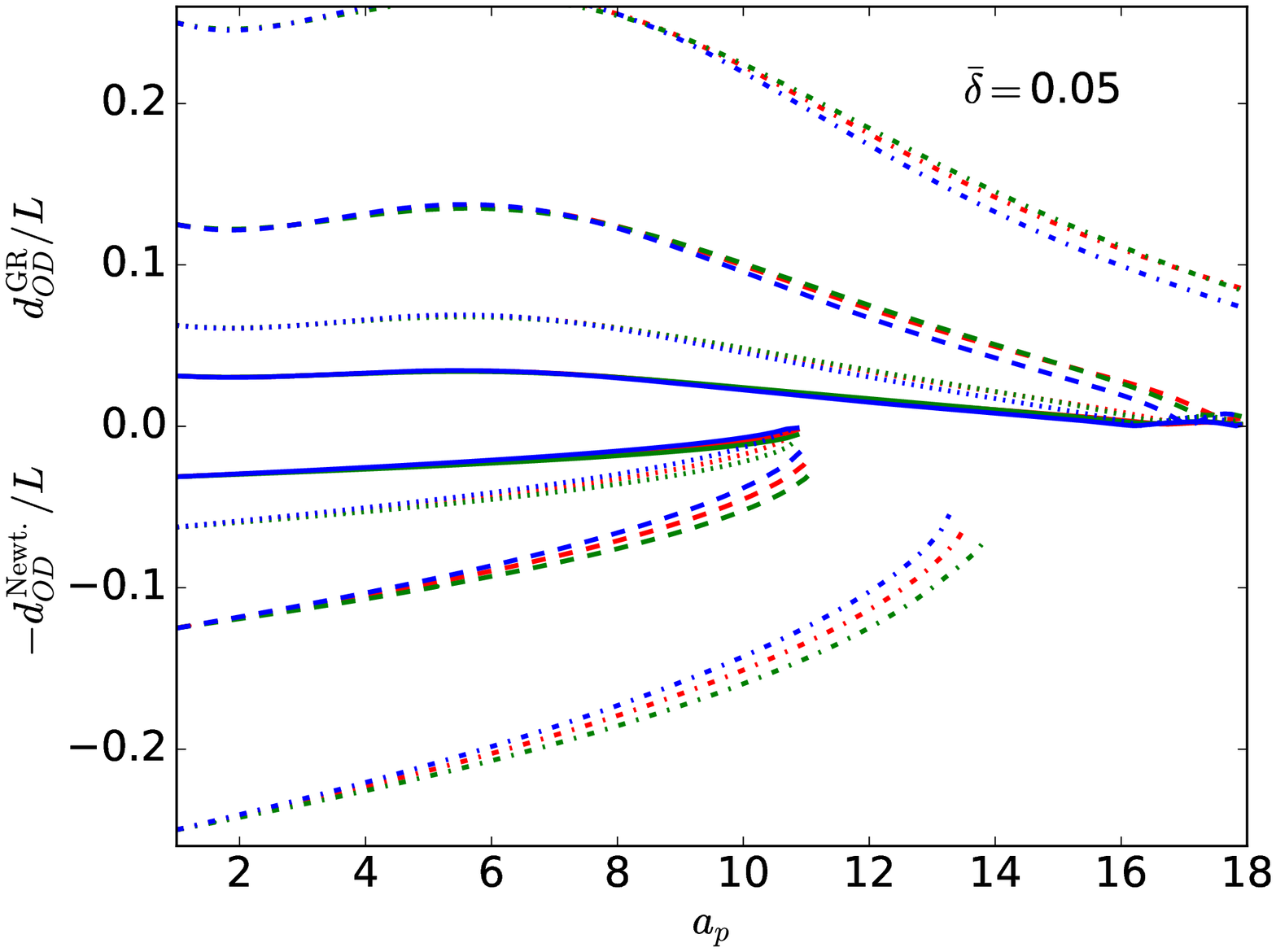}
\includegraphics[width=\columnwidth,draft=false]{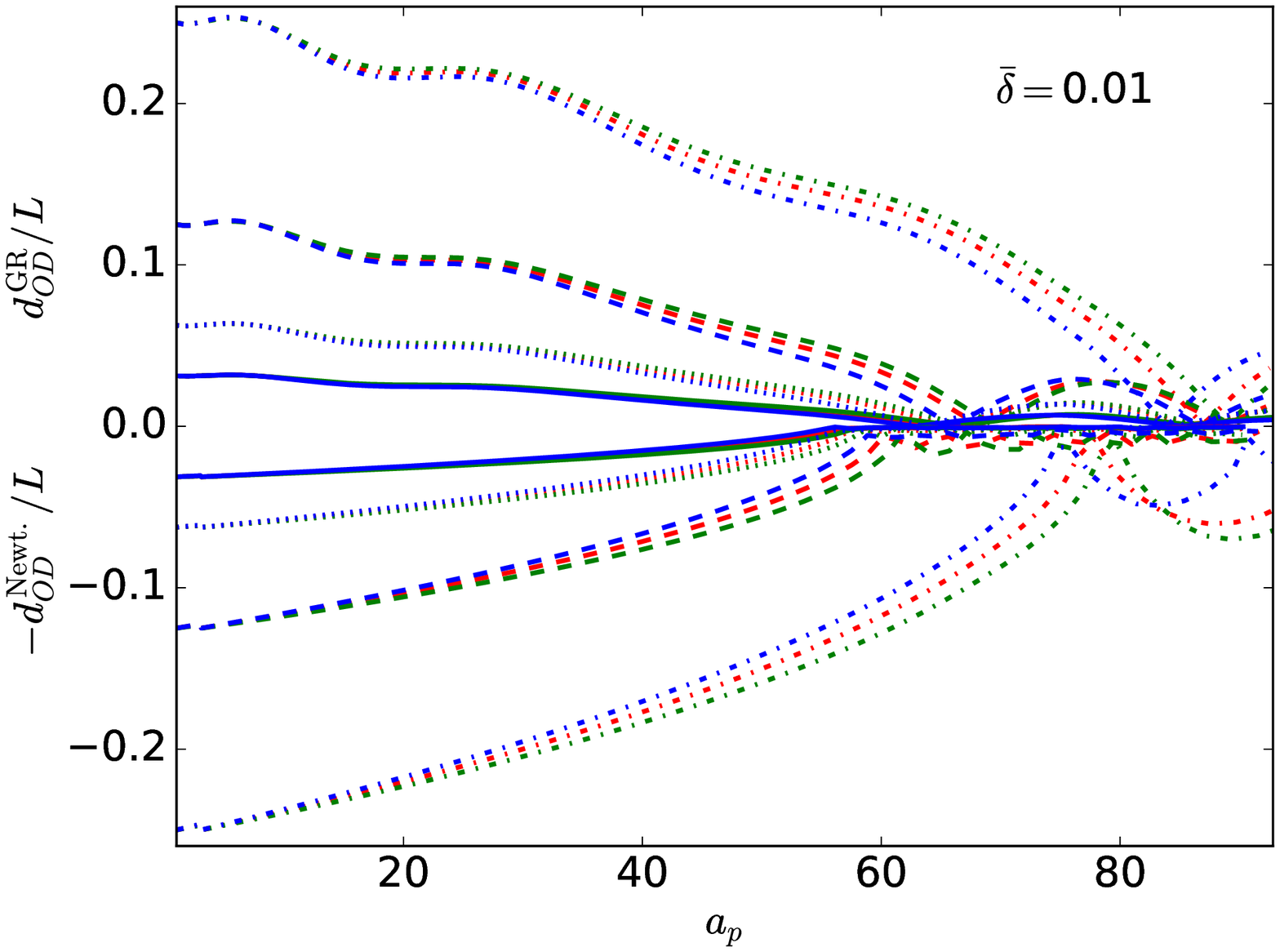}
\includegraphics[width=\columnwidth,draft=false]{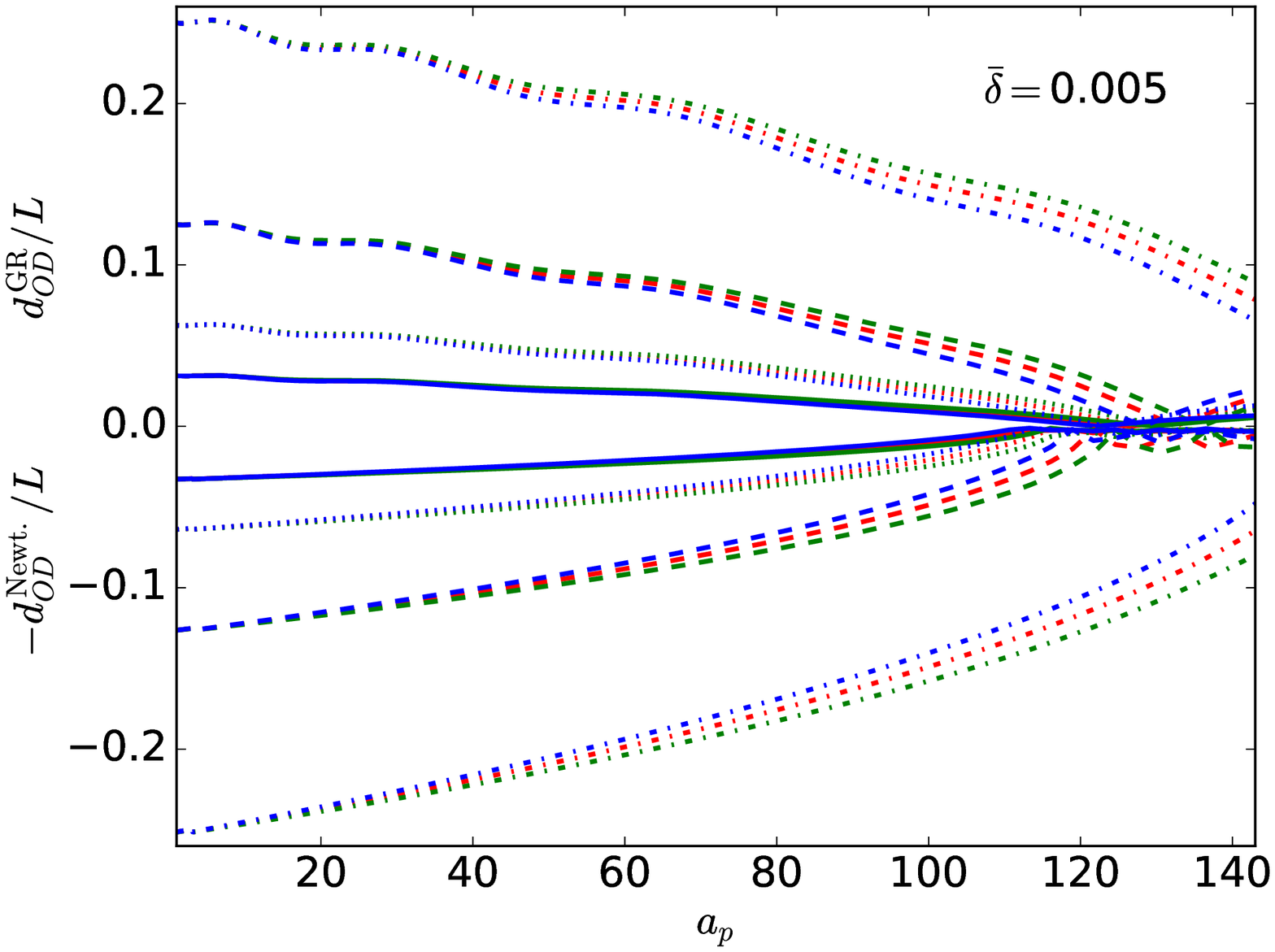}
\includegraphics[width=\columnwidth,draft=false]{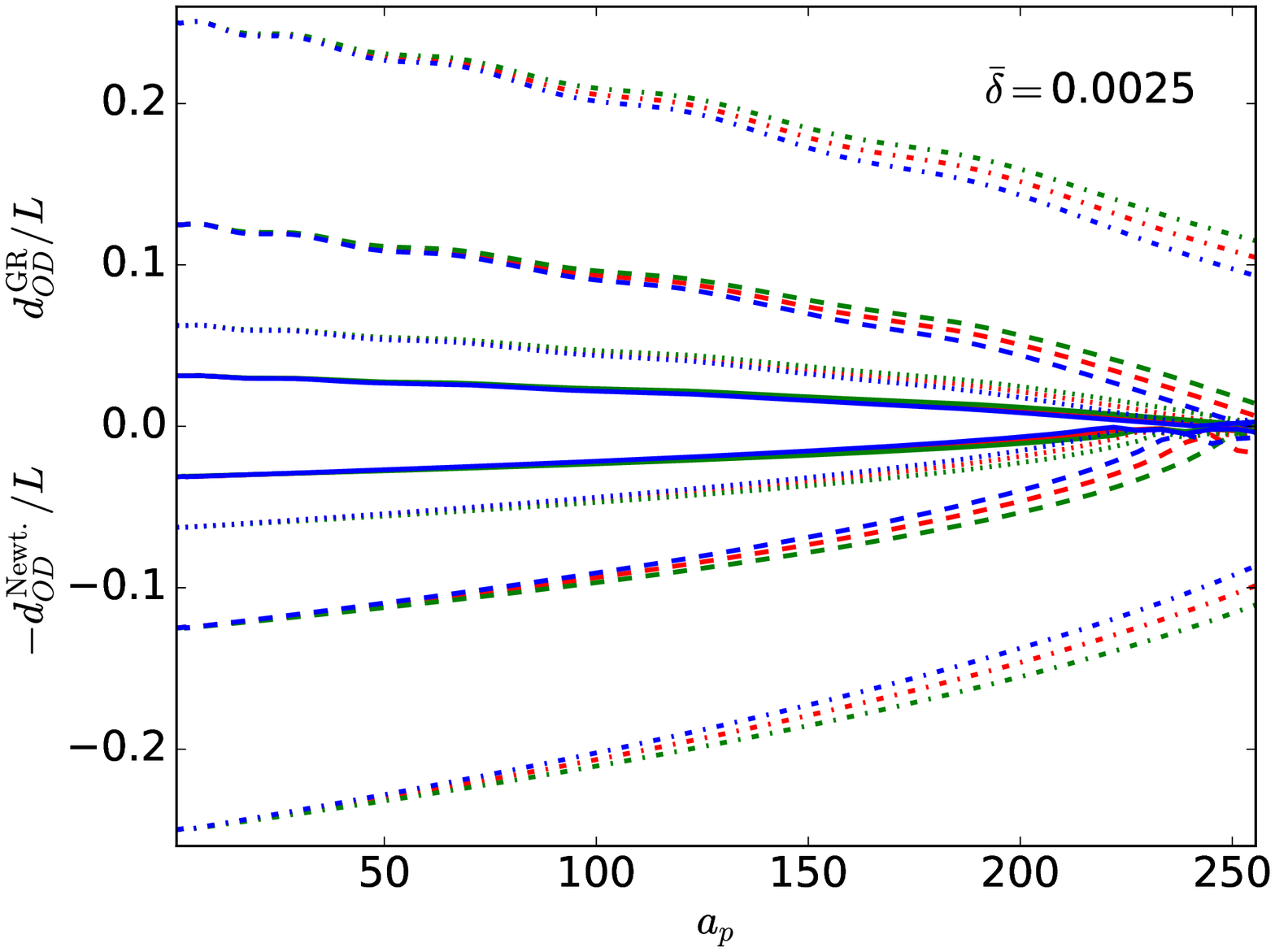}
\end{center}
\caption{
    The coordinate distances from the point of maximum density of a set
    of fiducial particles for the GR and Newtonian simulations,
    as a function of the proper-time scale factor of the particle.
    For each panel, the top half shows the GR results, while the bottom half shows
    the Newtonian results. The red, green, and blue curves correspond to particles
    initially displaced from the point of maximum overdensity in the x, y, and z coordinate
    directions, respectively.
    The different panels correspond to (left to right, top to bottom) $\bar{\delta}=0.05$,
    0.01, 0.005, and 0.0025.
    Though the actual distance is gauge dependent (which in particular is the reason
    for the initial oscillations in the GR curves), the time the particles cross
    the overdensity is not.
    For the $\bar{\delta}=0.05$ case, the Newtonian
calculation has to be terminated when the Newtonian potential becomes large.
\label{fig:splash}
}
\end{figure*}

The differences in the evolution of multistream regions can be tracked by considering a set
of fiducial observers, comoving with the matter, that are initially displaced
from the halo by some distance, and comparing the proper time it takes for them
to eventually fall through the point of maximum overdensity and begin
to oscillate around it.  This is
illustrated in Fig.~\ref{fig:splash}, where it is apparent that as the size of
the inhomogeneities increases, and the collapse takes place more quickly and at
scales more comparable to the Hubble scale, the relative discrepancy between
the Newtonian and GR cases increases, with the Newtonian case exhibiting faster
collapse.  (We note that in general these coordinate distances are gauge
dependent, but the time the particles cross the overdensity is not.)

\begin{figure}
\begin{center}
\includegraphics[width=\columnwidth,draft=false]{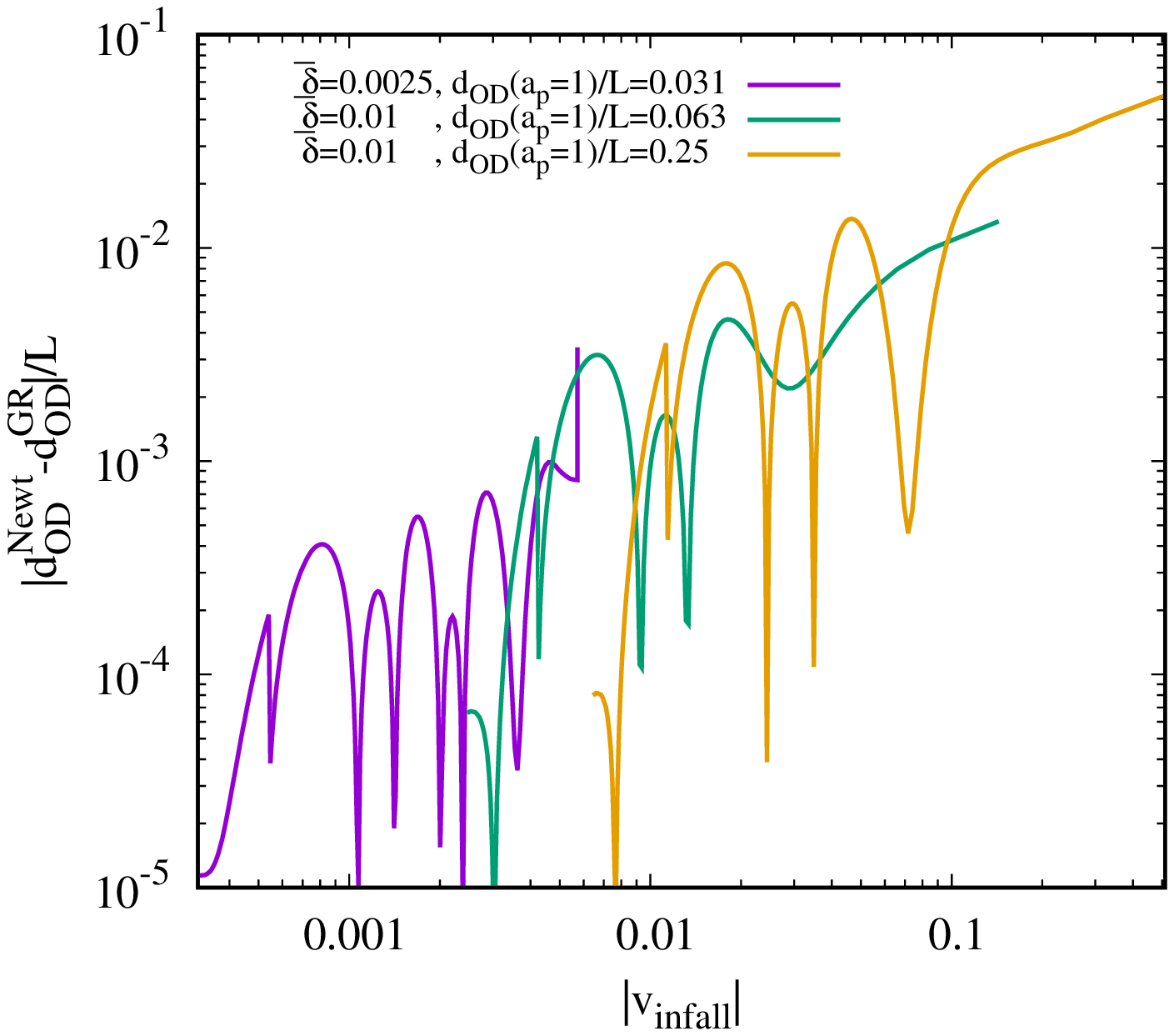}
\end{center}
\caption{
Differences in the coordinate distances between the GR and Newtonian simulations
    for a subset of freely falling fiducial particles from
    Fig.~\ref{fig:splash}, as a function of the absolute magnitude of infall
    velocity inferred from the Newtonian simulations. The Newtonian trajectories
    follow their GR counterparts quite closely , as long as the evolution is
    nonrelativistic. Significant differences between the 
    simulations occur when the evolution enters the relativistic regime.
  \label{fig:relativity}
}
\end{figure}

Figure~\ref{fig:relativity} shows the differences between the Newtonian and GR
positions of freely falling particles from Fig.~\ref{fig:splash} as a function
of the absolute magnitude of the infall velocity inferred from the Newtonian
simulation. For the sake of clarity, we only show the trajectories up until the
time where they first cross the halo center in the Newtonian run.  The
comparison demonstrates that the Newtonian trajectories closely follow their GR
counterparts, as long as infall velocities do not exceed the limits of
nonrelativistic dynamics. Noticeable discrepancies between the two simulations
occur when the particles reach relativistic velocities.  The apparent differences
reflect the limited accuracy of the Newtonian simulations when there is a violation of the
nonrelativistic assumption. Particles in the Newtonian simulations are
accelerated to larger velocities, giving rise to a faster collapse onto the
central object than in the GR simulations.

\begin{figure*}
\begin{center}
    \includegraphics[width=\columnwidth,draft=false]{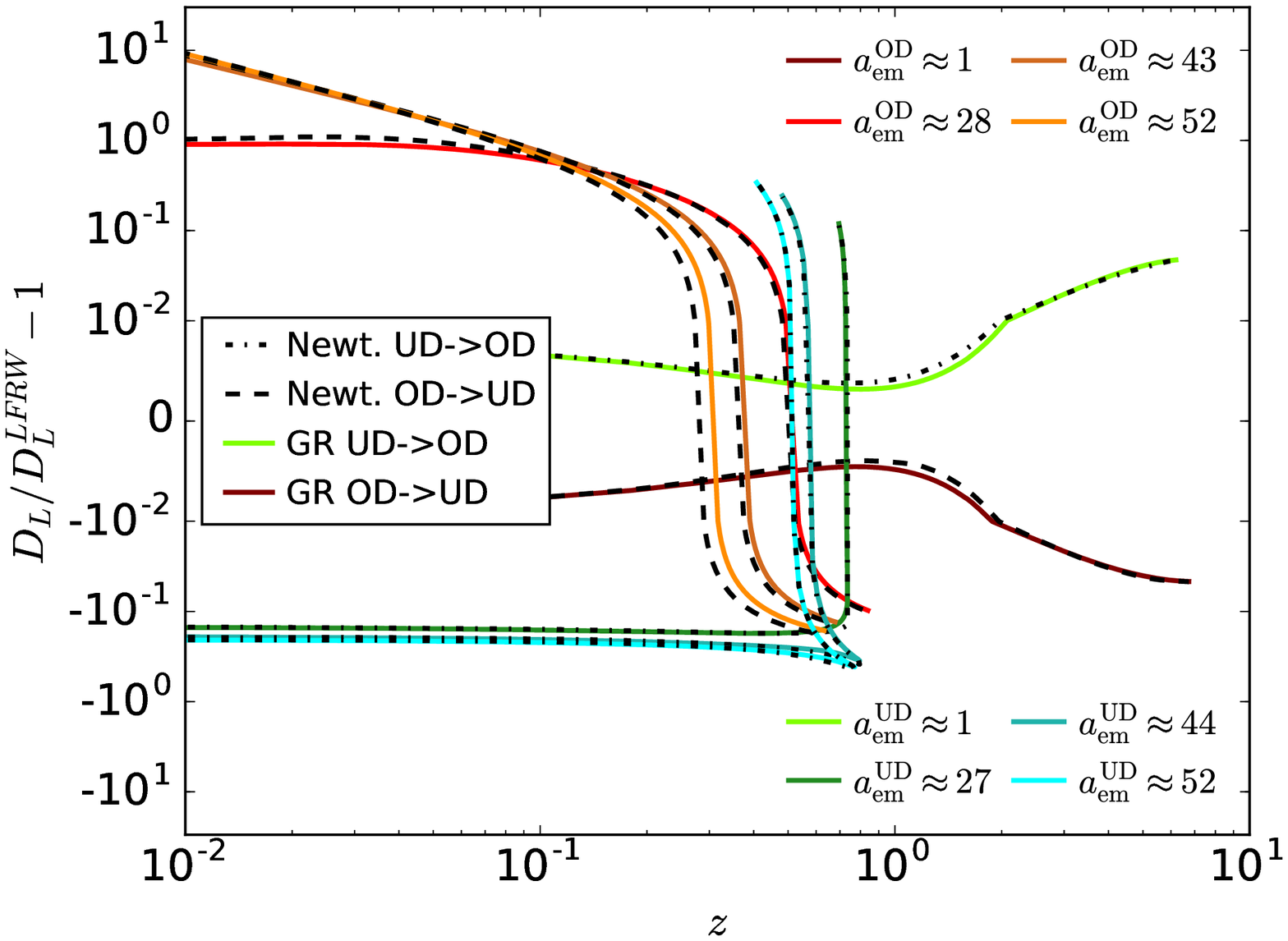}
    \includegraphics[width=\columnwidth,draft=false]{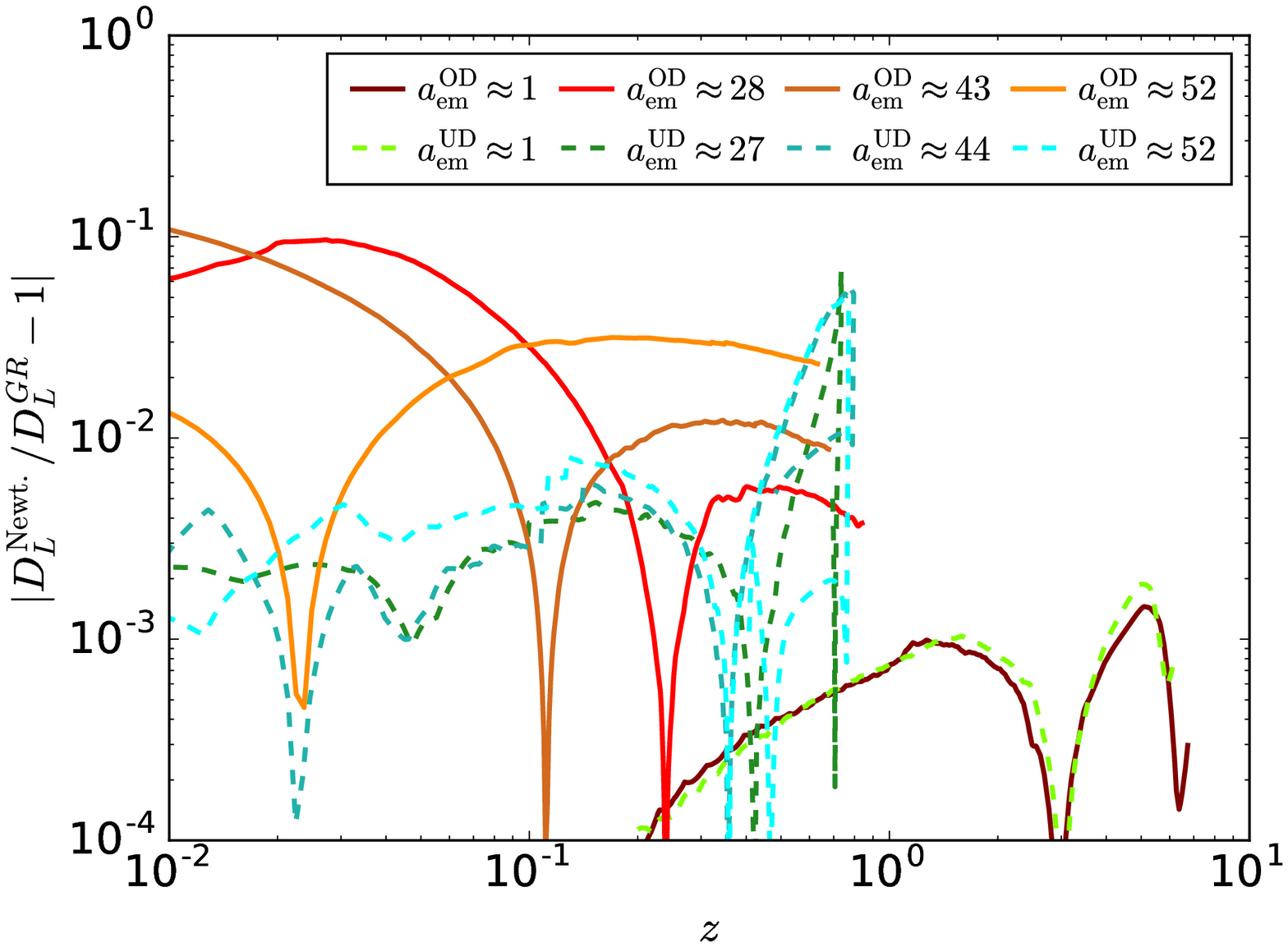}
    \includegraphics[width=\columnwidth,draft=false]{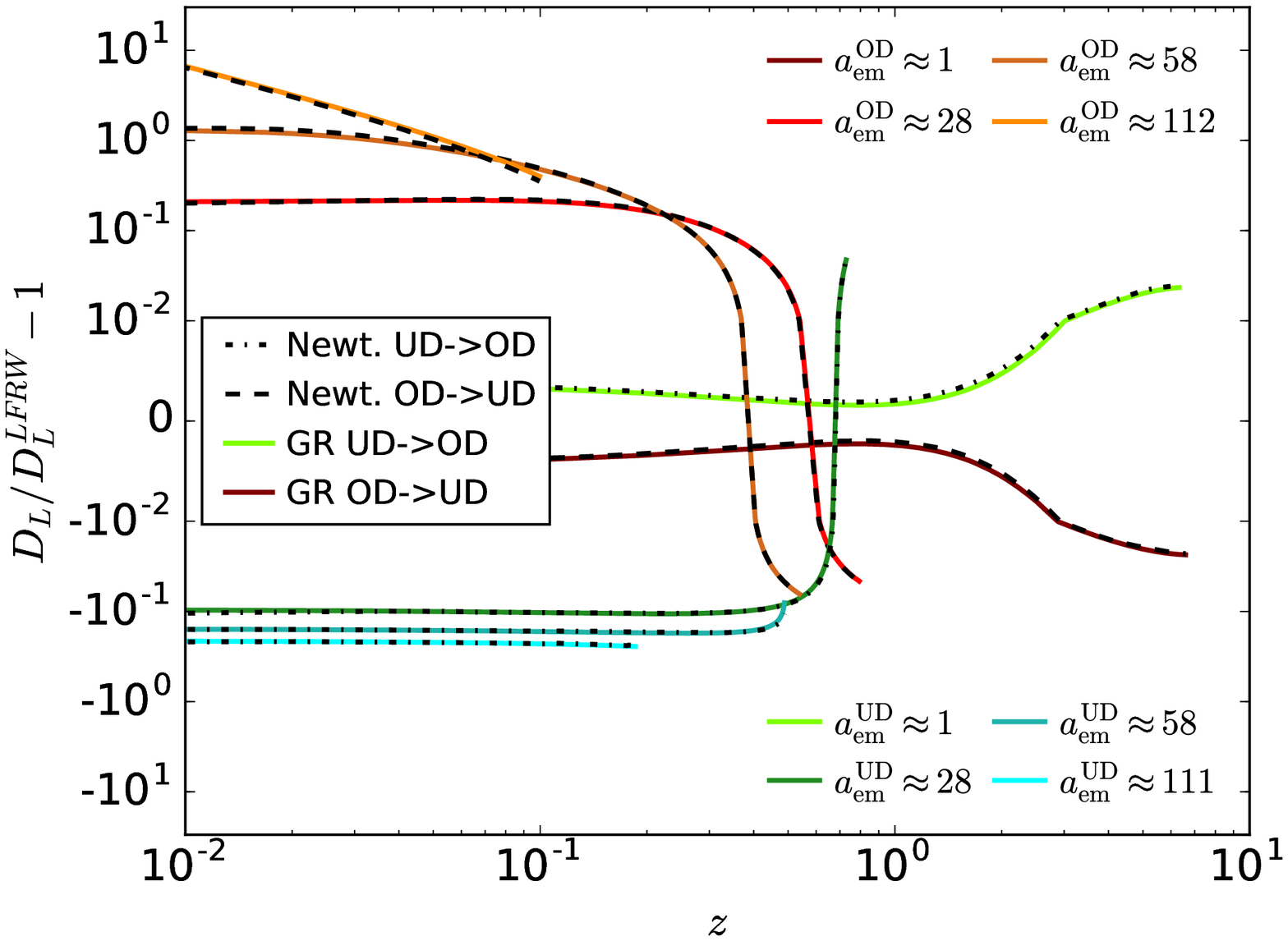}
    \includegraphics[width=\columnwidth,draft=false]{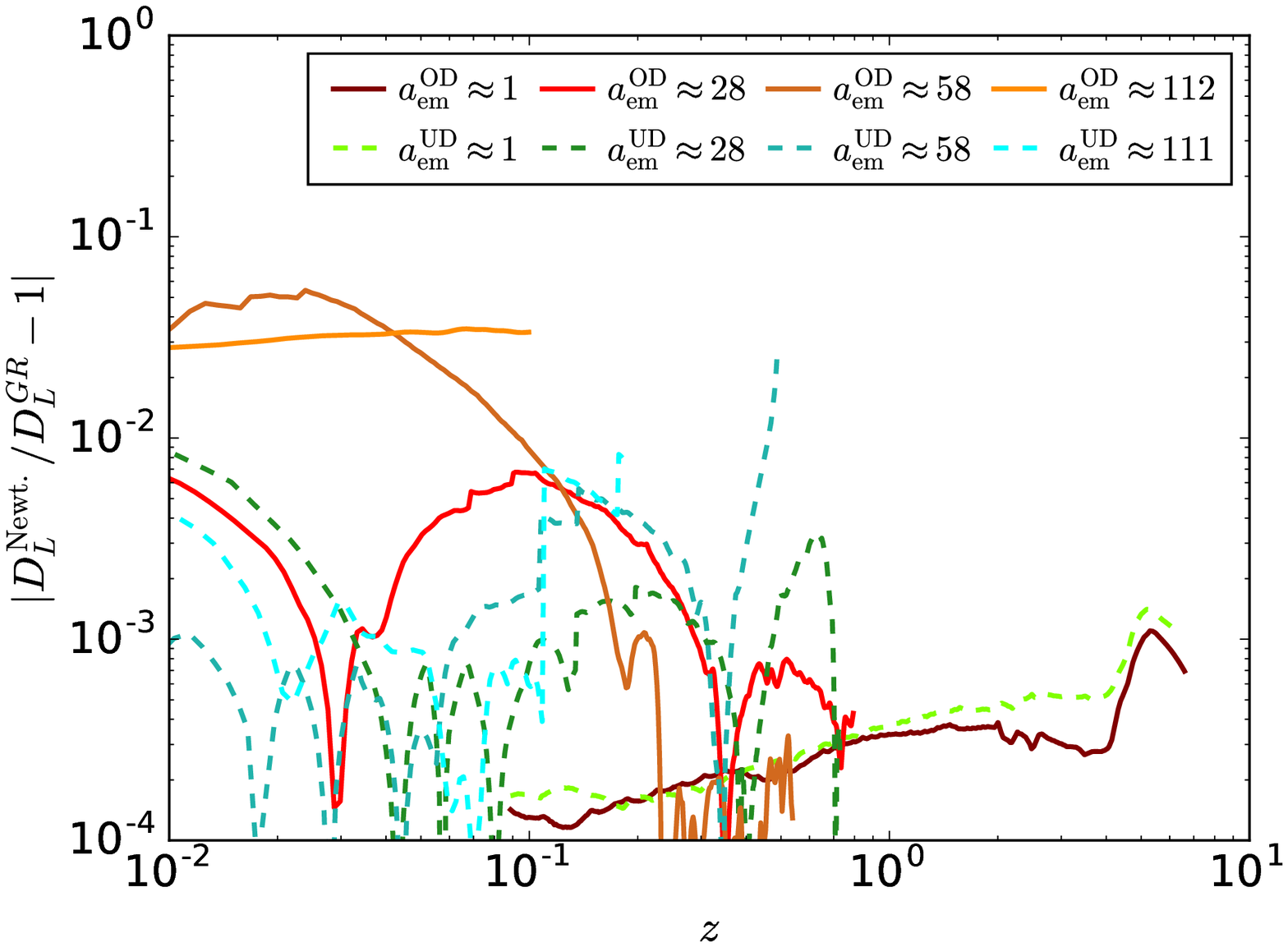}
    \includegraphics[width=\columnwidth,draft=false]{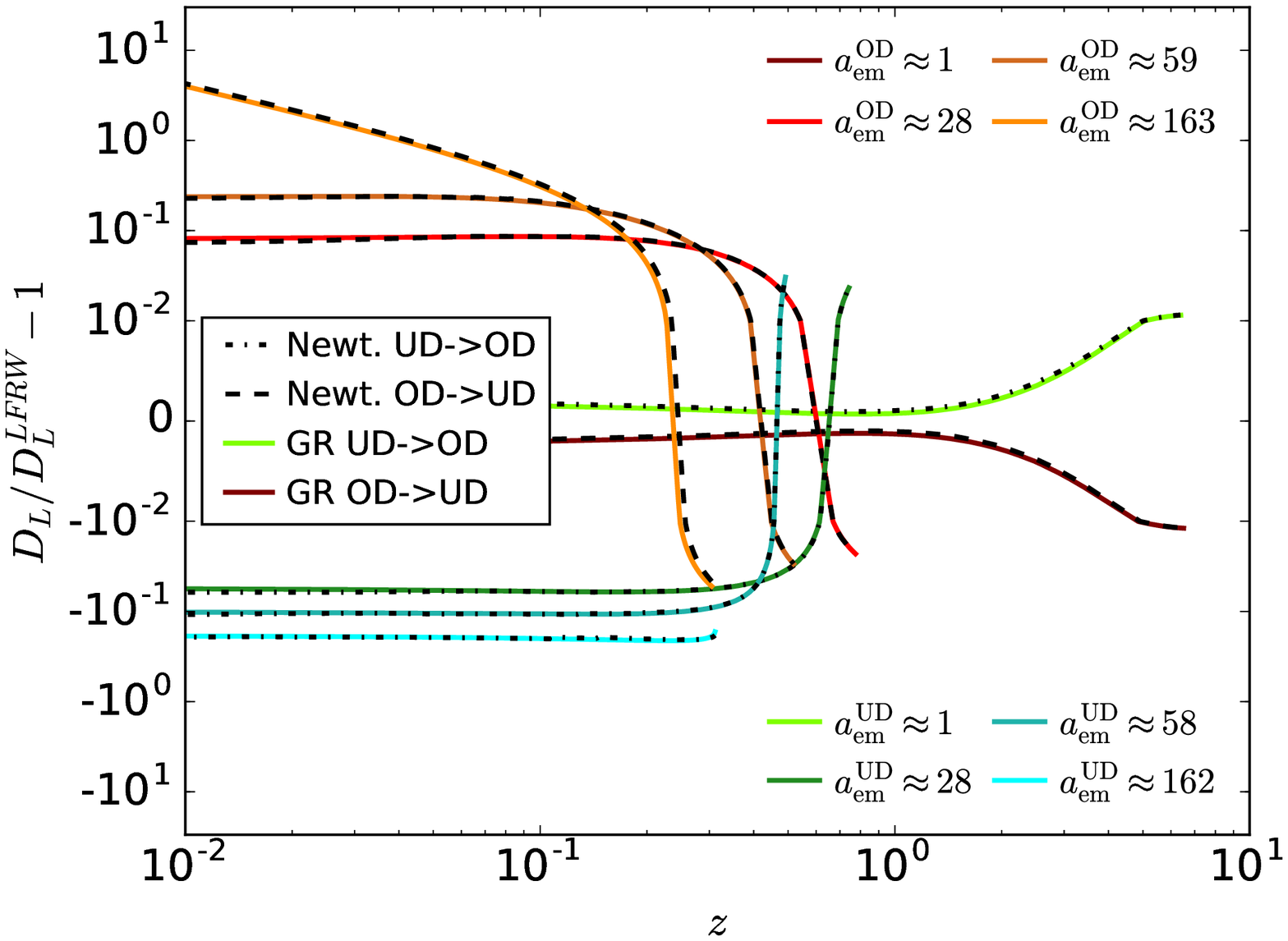}
    \includegraphics[width=\columnwidth,draft=false]{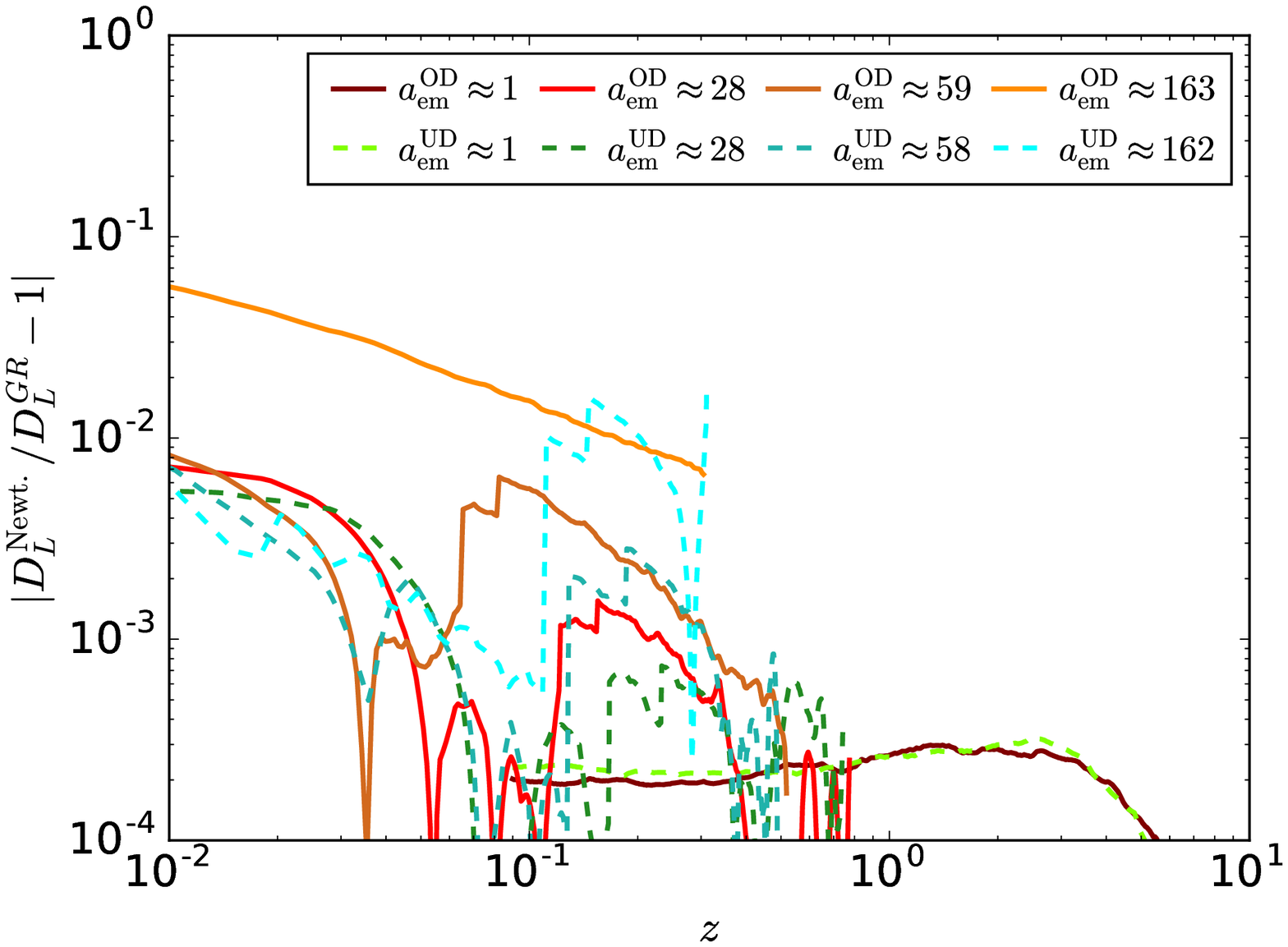}
\end{center}
\caption{
    The fractional difference in the luminosity distance versus redshift factor
    $D_L(z)$ for either the Newtonian or GR \N-body calculations from a
    homogeneous solution (left column), and from each other (right column), for
    a set of fiducial null geodesics that are emitted at the point of maximum
    density in the direction of the point of minimum density, or vice versa.
    Top to bottom, the different rows correspond to $\bar{\delta}=0.01$, 0.005, and 0.0025 . 
    In the left column, the vertical axis is linear from $-10^{-2}$ to $10^{-2}$, and
    logarithmic outside this range.
    We note that $z$ is defined individually for each null ray based on its emission
    time through Eq.~\ref{eqn:z}, as opposed to being a global quantity.
\label{fig:dl}
}
\end{figure*}

We can also compare the differences in the effective spacetimes using the
propagation of light.  In Fig.~\ref{fig:dl}, we compare the
luminosity--redshift relation for fiducial light rays propagating between the
points of minimum and maximum density.  From the comparison with the
homogeneous solution shown in the left column of Fig.~\ref{fig:dl}, one can see
that the cases considered here have large, nonlinear deviations from the LFRW
behavior.  Nevertheless, as evident in the right column, the differences
between the GR and Newtonian case remain much smaller, in most cases subpercent
and consistent with numerical truncation error (see the appendix and
Ref.~\cite{East:2017qmk}), indicating the differences in the spacetimes are
small.

For the larger amplitude inhomogeneities, light rays emitted from the
overdensity at later times have a $D_L(z)$ that is slightly smaller for the GR
calculation than the Newtonian counterpart at small $z$, but slightly larger
at larger $z$ as they move away from region of high gravitational potential.
For light rays emitted from the minimum density void, the differences between
the GR and Newtonian calculations generally remain small---at the subpercent
level---until the overdensity is approached. In the vicinity of the overdensity,
the gravitational potential can be strong enough to cause a blue-shift, as
evident in the top panel of Fig.~\ref{fig:dl}.

\begin{figure}
\begin{center}
\includegraphics[width=\columnwidth,draft=false]{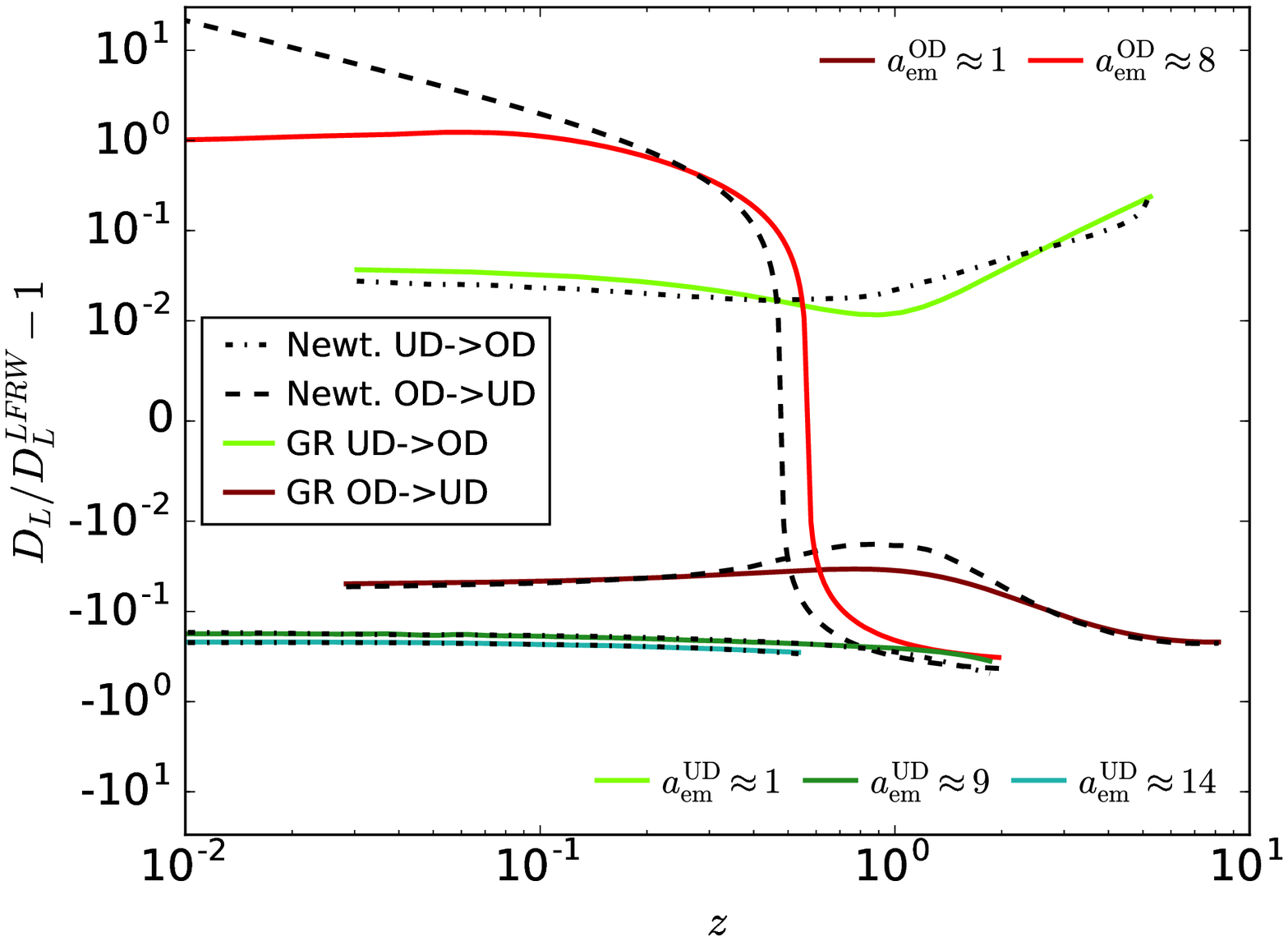}
\end{center}
\caption{
  Results for the highest amplitude perturbation case with $\bar{\delta}=0.05$,
    showing the fractional difference in the luminosity distance versus
    redshift factor $D_L(z)$ for either the Newtonian or GR \N-body
    calculations from a homogeneous solution, as in the left column of
    Fig.~\ref{fig:dl}.
  \label{fig:extreme}
}
\end{figure}

Finally, we mention further details of the case with $\bar{\delta}=0.05$. This
choice of initial conditions represents the extreme limiting case where the
Newtonian treatment completely breaks down, and the Newtonian potential reaches
$|\psi|\sim1/2$ after a 15-fold increase of scale factor.  
As shown in Fig.~\ref{fig:od_ud}, 
though the collapse at the
overdensity (middle panel) occurs faster (in terms of proper observer time) in the
Newtonian calculation than the full GR one, and the two calculations begin
to noticeably differ well before halo formation,
the evolution of the density in the void (top panel) still agrees well, with very little
``backreaction" of the high-curvature region on the global expansion.  In 
Fig.~\ref{fig:extreme}, we also show the luminosity
distance-redshift relation for this case, which continues the trend found in
Fig.~\ref{fig:dl}, with increasing deviation between the Newtonian and GR
calculations. Again, even for this extreme case, the differences between the
light propagation in the void region are small.  We are also not able to continue
the GR calculation forward indefinitely, but it appears that a black hole is
being formed at the overdensity. However, accurately tracking the attendant
small scales requires adaptive mesh refinement, which we leave to future work. 

\section{Discussion and Conclusion}
\label{sec:conclusion}
In this work, we have shown that a meaningful comparison can be carried out between
standard \N-body simulations of cosmological structure formation, which assume
Newtonian-type gravity on the background of a homogeneously expanding universe,
and full solutions of the Einstein-Vlasov equations, which make no assumptions
regarding a background cosmology.  For computational expediency, we have
focused on a simple set of initial conditions, with inhomogeneities at a single
length scale, but considered a range of amplitudes, including going all the way
to the limit where the nonrelativistic assumptions underlying the Newtonian
calculation break down. Tackling a more realistic power spectrum of density
fluctuations will require more advanced techniques, such as adaptive 
mesh refinement, and will be quite computationally expensive given the stringent 
requirements place on time steps due to the fact that information propagates
at the speed of light.

We find that for small initial density fluctuations, the Newtonian and GR
calculations show excellent agreement (with differences typically
subpercent and consistent with truncation error) well into the regime where the deviations
from homogeneity become nonlinear.  For large density fluctuations, the
dominant relativistic correction seems to be that the collapse of
overdensities occurs slower in the full GR calculation compared to the
Newtonian one. These discrepancies can already be anticipated from the
Newtonian calculation alone as the gravitational potential and infall
velocities are approaching relativistic values. Even for such cases, the effect
on the expansion outside the high density/velocity regions (e.g. in the voids)
is found to be small, bounding backreaction effects.

Comparing the properties of light propagation in the Newtonian and GR
calculations, we demonstrated that the resulting distance-redshift relations
agree at the subpercent level as long as the Newtonian potential does not
exceed the limit of a weak field approximation, i.e. $|\Psi_{N}|\leq0.1$.  As a
limiting case, we have considered initial conditions all the way up to ones where
the fluctuations in the density exceed the average value at the corresponding
scales in the standard $\Lambda$CDM model by factor of $\sim500$ (that is, at
the present time they roughly correspond to $\sim 0.5$ at a Gpc scale).  Since
our simulations test the evolution on cosmological scales of perturbations with
amplitudes exceeding those applicable to observational cosmology, we conclude
that the obtained results provide a strong validation of the standard Newtonian
approach employed in observational cosmology.  In particular, our comparison
implies that GR corrections to the Newtonian calculation of the cosmic variance
in the local measurement of the Hubble constant are negligible.  This
strengthens the conclusion that a $\sim9\%$ difference between the local and
cosmic microwave background (CMB) based measurements of the Hubble constant,
currently at $4.4\sigma$ statistical significance \citep{Rie2019}, cannot be
ascribed to the cosmic variance which is estimated at $\sim0.5$ percent
\citep{Wojtak2014,Wu2017,Odde2014}.
This in line with the conclusion of a recent study in Ref.~\cite{Macpherson2018} that
looked at variations in the local expansion in a particular gauge  
using GR-fluid simulations (that hence cannot describe multistream regions)
with a cosmologically motivated power spectrum. 

The methods described here could be applied to study the formation of
primordial black holes during a matter-dominated era (see
e.g.~\cite{Carr:2017edp} and references therein), or scenarios where black
holes make up some fraction of the dark matter.  They could also be used to
study ultralarge scale structure~\cite{Braden:2016tjn}, which could be related to
understanding persistent CMB anomalies at large angular scales, which seem to
indicate a violation of statistical isotropy and scale invariance of
inflationary perturbations \citep{Schwarz2016}. Comparable scales will be also
probed by the upcoming deep imaging cosmological surveys. In particular, the
Large Synoptic Survey Telescope will reach an unprecedented effective volume of
$\sim 4 H_{0}^{-3}$ \citep{LSST2009}.

\acknowledgments
W.E.E. acknowledges support from an NSERC Discovery grant.
This research was supported in part
by Perimeter Institute for Theoretical Physics.  Research at Perimeter
Institute is supported by the Government of Canada through the Department of
Innovation, Science and Economic Development Canada and by the Province of
Ontario through the Ministry of Research, Innovation and Science. R.W. 
was supported by a grant from VILLUM FONDEN (Project No. 16599).
F.P.  acknowledges support from NSF grant PHY-1912171, the Simons
Foundation, and the Canadian Institute For Advanced Research (CIFAR).
Computational resources were provided by XSEDE under grant
TG-PHY100053 and the Perseus cluster at Princeton University.

\appendix
\section*{Appendix: Numerical error results}
\begin{figure}
\begin{center}
\includegraphics[width=\columnwidth,draft=false]{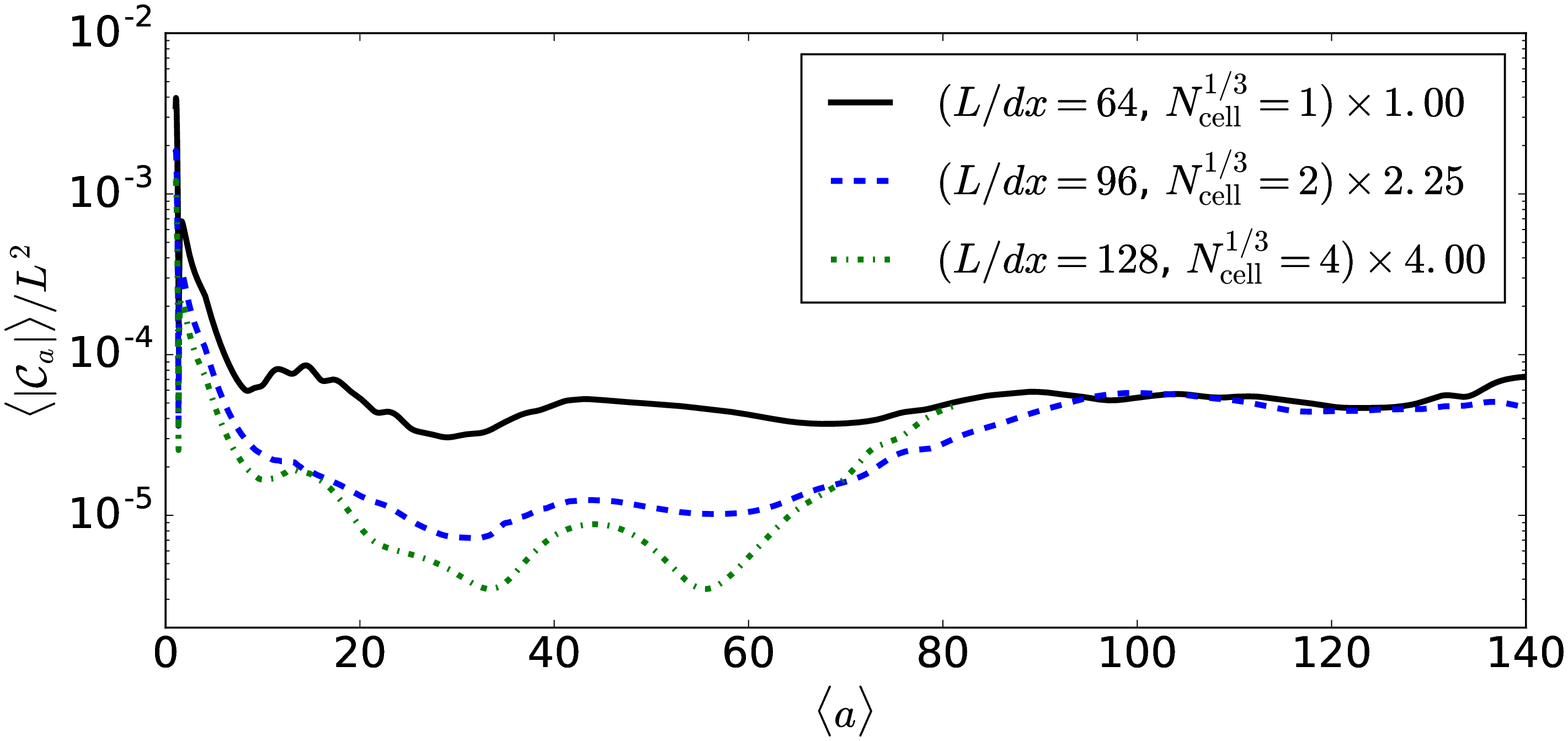}
\end{center}
\caption{
    Convergence of the L2 norm of the generalized harmonic constraint
    ($C_a:=H_a-\Box x_a$) for the $\bar{\delta}=10^{-2}$ case, shown as a
    function of a volume-averaged measure of the scale factor. The different
    resolutions have been scaled assuming second order convergence with the
    grid spacing, though at later times error from the finite number of
    particles begins to dominate.
\label{fig:cnst_conv}
}
\end{figure}

In this appendix, we include some results on numerical convergence.  For the GR
simulations, we initially find the numerical error to be dominated by the grid
spacing, which also sets the integration time step. However, at late times, as
large under and overdensities develop, the number of particles used to sample
the matter distribution becomes important.  In Fig.~\ref{fig:cnst_conv}, we
show the convergence of the Einstein constraints with increasing numerical
resolution for the $\bar{\delta}=10^{-2}$ case.  The results have been scaled
assuming second order convergence with grid spacing.

\begin{figure}
\begin{center}
\includegraphics[width=\columnwidth,draft=false]{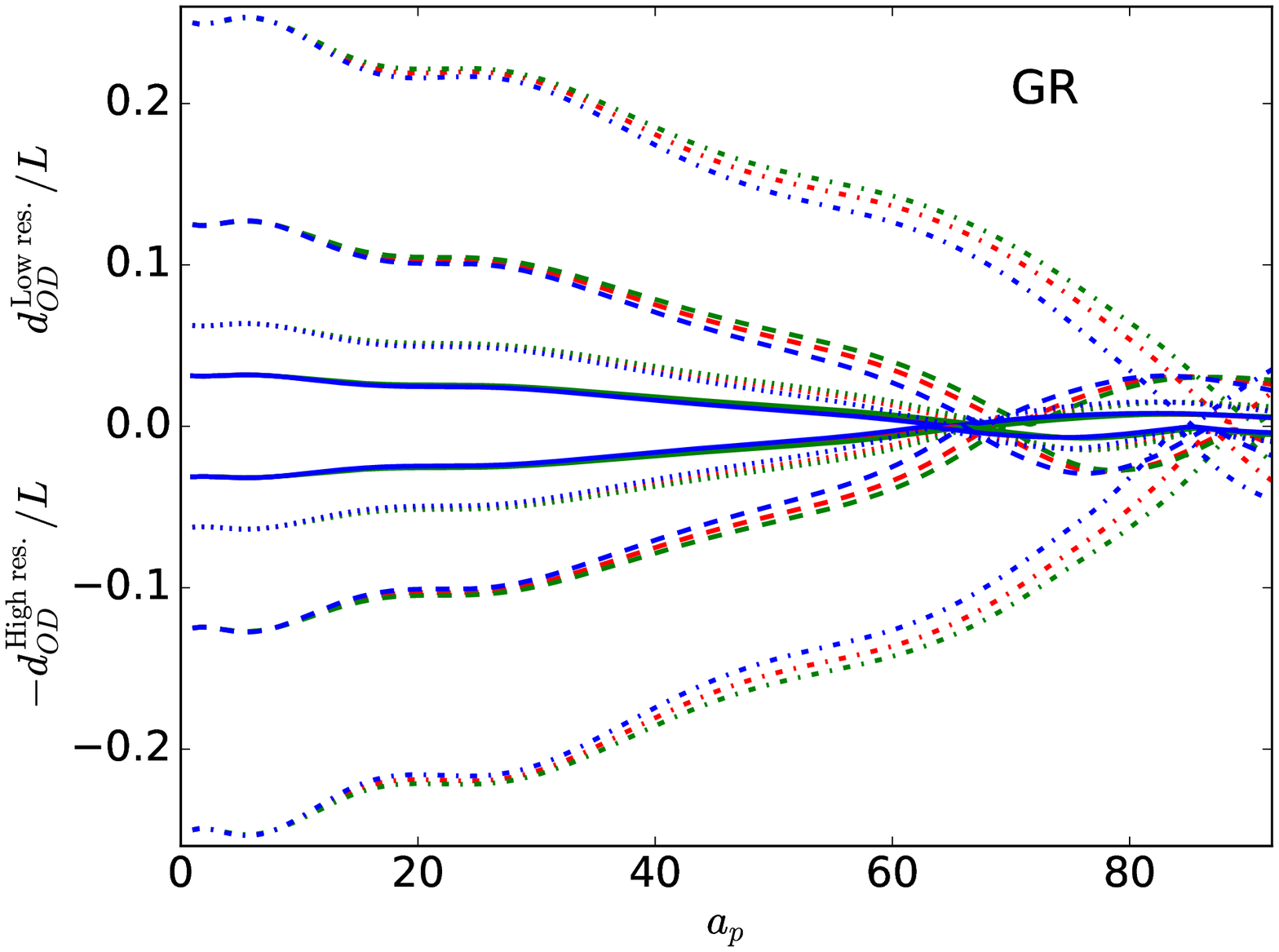}
    \includegraphics[width=\columnwidth,draft=false]{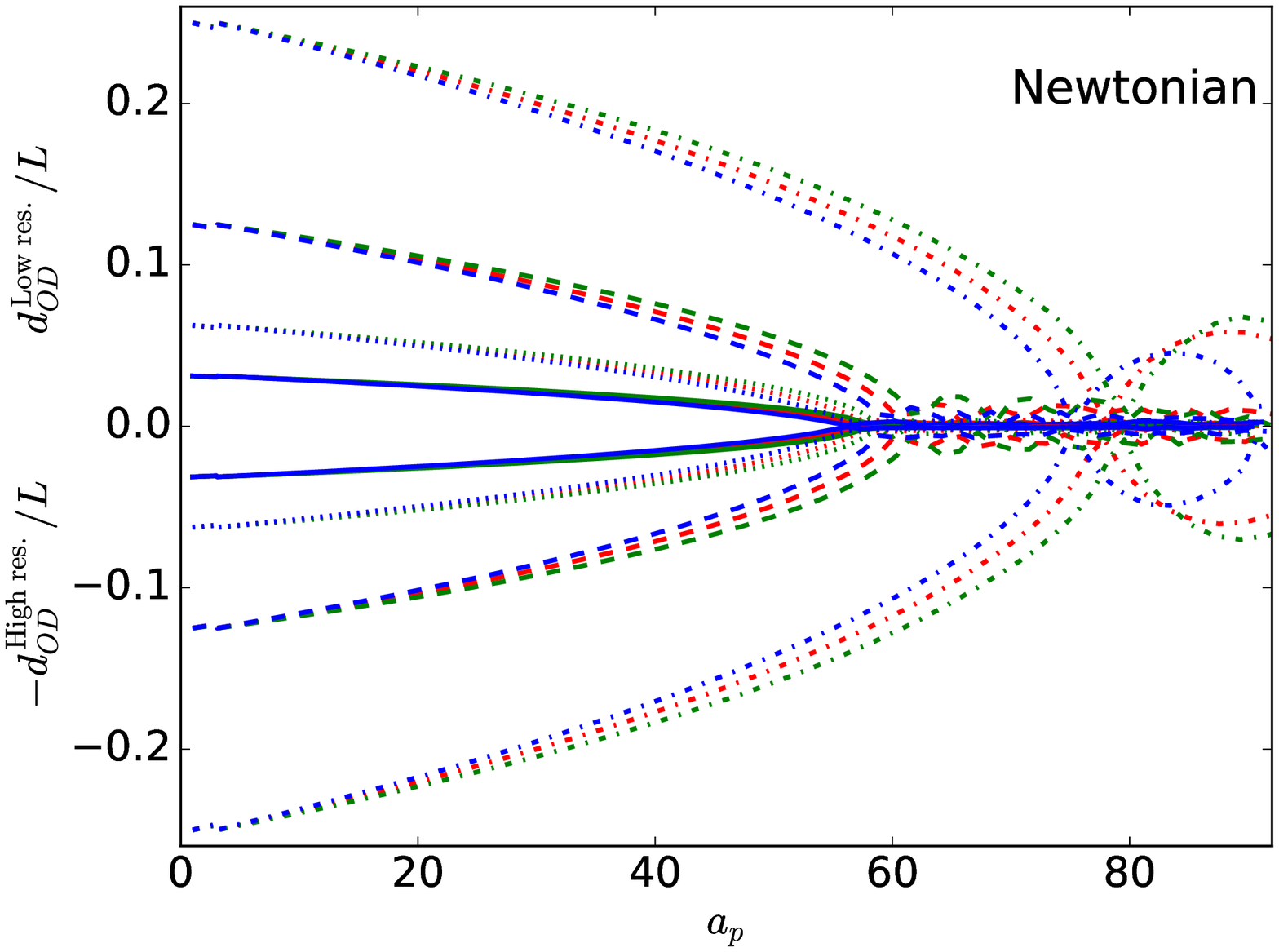}
\end{center}
\caption{
    Similar to the top right panel of Fig.~\ref{fig:splash} ($\bar{\delta}=0.01$),
    but showing the dependence of resolution. The top panel shows 
    GR simulations with $N=128^3$ (top half of panel) and $N=384^3$ (bottom half of panel) number of particles, and
    the bottom panel shows Newtonian simulations with $N=128^3$ (top half) and $N=196^3$ (bottom half).
\label{fig:splash_conv}
}
\end{figure}

In Fig.~\ref{fig:splash_conv}, we show the halo crossing time for this same case
as a function of resolution, for both the GR and Newtonian simulations.
The discrepancies with resolution in the time of first crossing are small
compared to the differences between the GR and Newtonian simulations 
(though they do become more pronounced for subsequent oscillations).

\begin{figure}
\begin{center}
    \includegraphics[width=\columnwidth,draft=false]{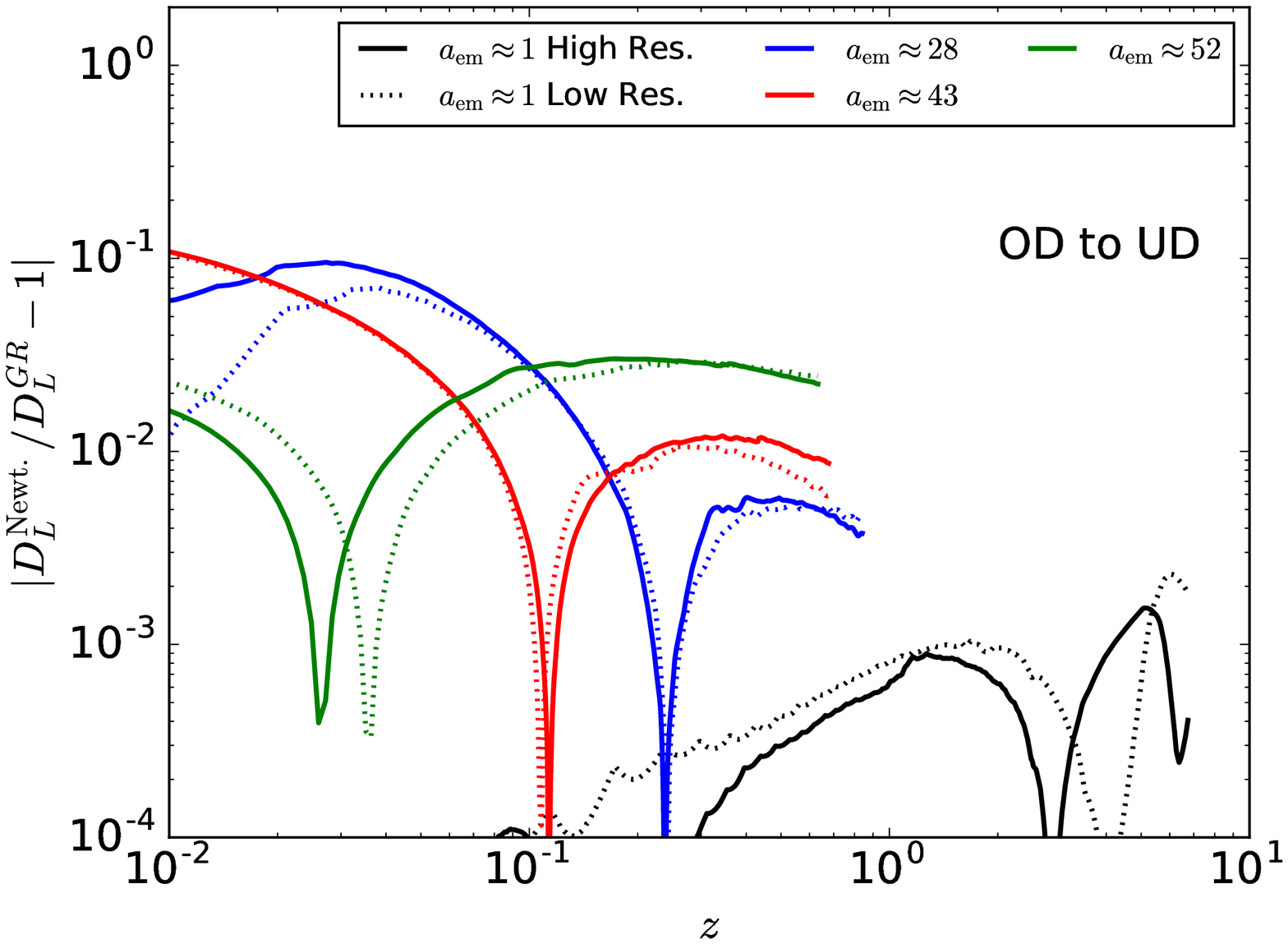}
    \includegraphics[width=\columnwidth,draft=false]{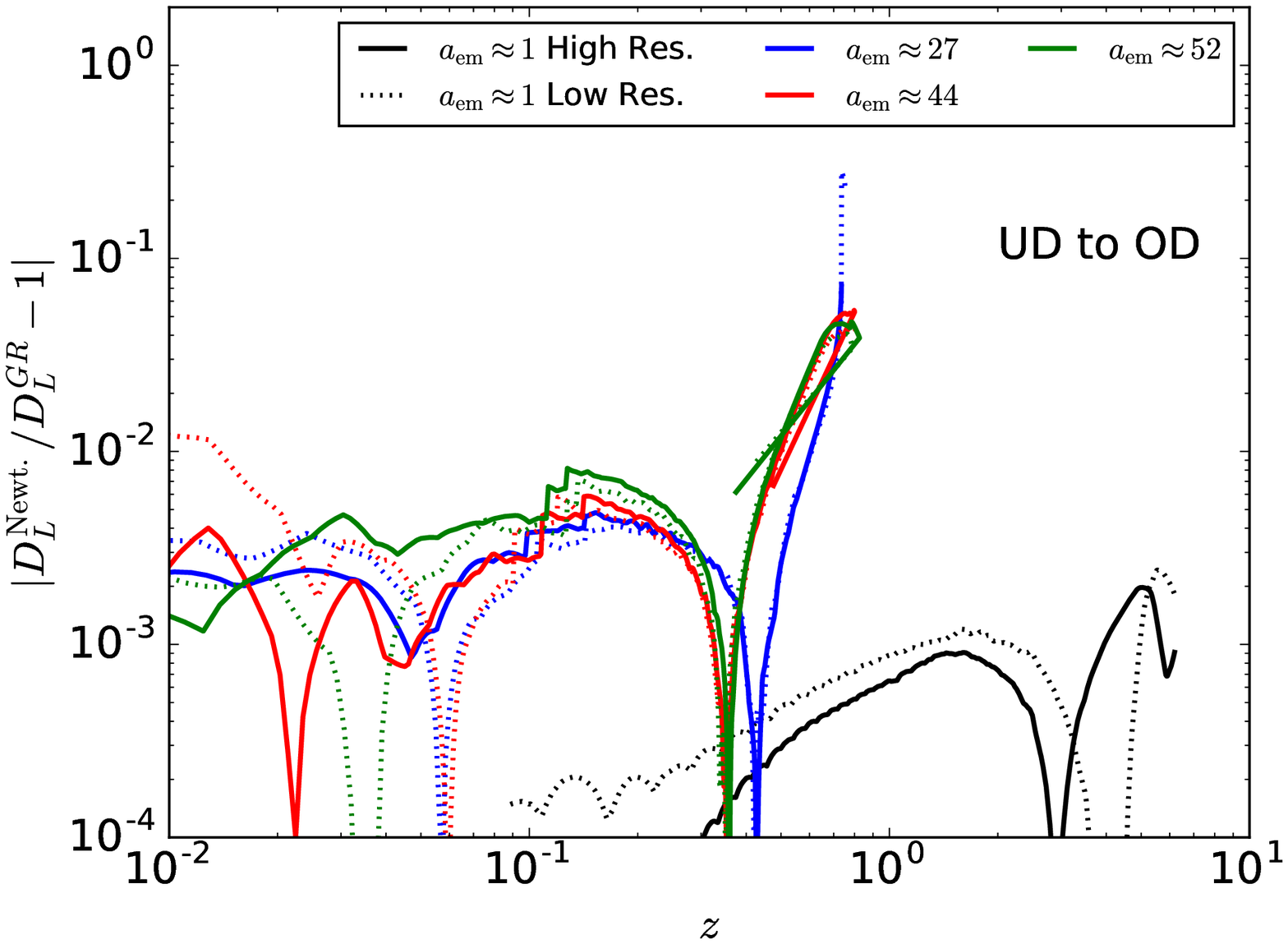}
\end{center}
\caption{
    Similar to the top left panel of Fig.~\ref{fig:dl} ($\bar{\delta}=0.01$),
    but showing the dependence on resolution. 
    The low resolution results compare 
    GR simulations with $N=192^3$  particles to Newtonian simulations with $N=128^3$,
    while the high resolution results compare GR simulations with $N=512^3$ to Newtonian
    simulations with $N=196^3$.
    The top panel shows 
    the null geodesics emitted at the overdensity, while the bottom panel shows
    those emitted at the underdensities.
\label{fig:dl_conv}
}
\end{figure}

Finally, we compare the resolution dependence of the luminosity distance-redshift
measures in Fig.~\ref{fig:dl_conv}.
From this it can be seen that most of the $\lesssim 1\% $ differences between 
the GR and Newtonian simulations seen at early
times or in the propagation outside the very high density regime are attributable
just to truncation error. In contrast, the significant differences in propagation
in the vicinity of the large overdensity exceed the truncation error, and
in some cases are underestimated at lower resolutions.

\bibliographystyle{apsrev4-1}
\bibliography{ref}

\begin{thebibliography}{46}%
\makeatletter
\providecommand \@ifxundefined [1]{%
 \@ifx{#1\undefined}
}%
\providecommand \@ifnum [1]{%
 \ifnum #1\expandafter \@firstoftwo
 \else \expandafter \@secondoftwo
 \fi
}%
\providecommand \@ifx [1]{%
 \ifx #1\expandafter \@firstoftwo
 \else \expandafter \@secondoftwo
 \fi
}%
\providecommand \natexlab [1]{#1}%
\providecommand \enquote  [1]{``#1''}%
\providecommand \bibnamefont  [1]{#1}%
\providecommand \bibfnamefont [1]{#1}%
\providecommand \citenamefont [1]{#1}%
\providecommand \href@noop [0]{\@secondoftwo}%
\providecommand \href [0]{\begingroup \@sanitize@url \@href}%
\providecommand \@href[1]{\@@startlink{#1}\@@href}%
\providecommand \@@href[1]{\endgroup#1\@@endlink}%
\providecommand \@sanitize@url [0]{\catcode `\\12\catcode `\$12\catcode
  `\&12\catcode `\#12\catcode `\^12\catcode `\_12\catcode `\%12\relax}%
\providecommand \@@startlink[1]{}%
\providecommand \@@endlink[0]{}%
\providecommand \url  [0]{\begingroup\@sanitize@url \@url }%
\providecommand \@url [1]{\endgroup\@href {#1}{\urlprefix }}%
\providecommand \urlprefix  [0]{URL }%
\providecommand \Eprint [0]{\href }%
\providecommand \doibase [0]{http://dx.doi.org/}%
\providecommand \selectlanguage [0]{\@gobble}%
\providecommand \bibinfo  [0]{\@secondoftwo}%
\providecommand \bibfield  [0]{\@secondoftwo}%
\providecommand \translation [1]{[#1]}%
\providecommand \BibitemOpen [0]{}%
\providecommand \bibitemStop [0]{}%
\providecommand \bibitemNoStop [0]{.\EOS\space}%
\providecommand \EOS [0]{\spacefactor3000\relax}%
\providecommand \BibitemShut  [1]{\csname bibitem#1\endcsname}%
\let\auto@bib@innerbib\@empty
\bibitem [{\citenamefont {Bentivegna}\ and\ \citenamefont
  {Bruni}(2016)}]{Bentivegna:2015flc}%
  \BibitemOpen
  \bibfield  {author} {\bibinfo {author} {\bibfnamefont {E.}~\bibnamefont
  {Bentivegna}}\ and\ \bibinfo {author} {\bibfnamefont {M.}~\bibnamefont
  {Bruni}},\ }\href {\doibase 10.1103/PhysRevLett.116.251302} {\bibfield
  {journal} {\bibinfo  {journal} {Phys. Rev. Lett.}\ }\textbf {\bibinfo
  {volume} {116}},\ \bibinfo {pages} {251302} (\bibinfo {year} {2016})},\
  \Eprint {http://arxiv.org/abs/1511.05124} {arXiv:1511.05124 [gr-qc]}
  \BibitemShut {NoStop}%
\bibitem [{\citenamefont {Giblin}\ \emph
  {et~al.}(2016{\natexlab{a}})\citenamefont {Giblin}, \citenamefont {Mertens},\
  and\ \citenamefont {Starkman}}]{Giblin:2015vwq}%
  \BibitemOpen
  \bibfield  {author} {\bibinfo {author} {\bibfnamefont {J.~T.}\ \bibnamefont
  {Giblin}}, \bibinfo {author} {\bibfnamefont {J.~B.}\ \bibnamefont {Mertens}},
  \ and\ \bibinfo {author} {\bibfnamefont {G.~D.}\ \bibnamefont {Starkman}},\
  }\href {\doibase 10.1103/PhysRevLett.116.251301} {\bibfield  {journal}
  {\bibinfo  {journal} {Phys. Rev. Lett.}\ }\textbf {\bibinfo {volume} {116}},\
  \bibinfo {pages} {251301} (\bibinfo {year} {2016}{\natexlab{a}})},\ \Eprint
  {http://arxiv.org/abs/1511.01105} {arXiv:1511.01105 [gr-qc]} \BibitemShut
  {NoStop}%
\bibitem [{\citenamefont {Rekier}\ \emph {et~al.}(2015)\citenamefont {Rekier},
  \citenamefont {Cordero-Carrión},\ and\ \citenamefont
  {Füzfa}}]{Rekier:2014rqa}%
  \BibitemOpen
  \bibfield  {author} {\bibinfo {author} {\bibfnamefont {J.}~\bibnamefont
  {Rekier}}, \bibinfo {author} {\bibfnamefont {I.}~\bibnamefont
  {Cordero-Carrión}}, \ and\ \bibinfo {author} {\bibfnamefont
  {A.}~\bibnamefont {Füzfa}},\ }\href {\doibase 10.1103/PhysRevD.91.024025}
  {\bibfield  {journal} {\bibinfo  {journal} {Phys. Rev.}\ }\textbf {\bibinfo
  {volume} {D91}},\ \bibinfo {pages} {024025} (\bibinfo {year} {2015})},\
  \Eprint {http://arxiv.org/abs/1409.3476} {arXiv:1409.3476 [gr-qc]}
  \BibitemShut {NoStop}%
\bibitem [{\citenamefont {Giblin}\ \emph
  {et~al.}(2016{\natexlab{b}})\citenamefont {Giblin}, \citenamefont {Mertens},\
  and\ \citenamefont {Starkman}}]{Giblin:2016mjp}%
  \BibitemOpen
  \bibfield  {author} {\bibinfo {author} {\bibfnamefont {J.~T.}\ \bibnamefont
  {Giblin}}, \bibinfo {author} {\bibfnamefont {J.~B.}\ \bibnamefont {Mertens}},
  \ and\ \bibinfo {author} {\bibfnamefont {G.~D.}\ \bibnamefont {Starkman}},\
  }\href {\doibase 10.3847/1538-4357/833/2/247} {\bibfield  {journal} {\bibinfo
   {journal} {Astrophys. J.}\ }\textbf {\bibinfo {volume} {833}},\ \bibinfo
  {pages} {247} (\bibinfo {year} {2016}{\natexlab{b}})},\ \Eprint
  {http://arxiv.org/abs/1608.04403} {arXiv:1608.04403 [astro-ph.CO]}
  \BibitemShut {NoStop}%
\bibitem [{\citenamefont {Macpherson}\ \emph {et~al.}(2017)\citenamefont
  {Macpherson}, \citenamefont {Lasky},\ and\ \citenamefont
  {Price}}]{Macpherson:2016ict}%
  \BibitemOpen
  \bibfield  {author} {\bibinfo {author} {\bibfnamefont {H.~J.}\ \bibnamefont
  {Macpherson}}, \bibinfo {author} {\bibfnamefont {P.~D.}\ \bibnamefont
  {Lasky}}, \ and\ \bibinfo {author} {\bibfnamefont {D.~J.}\ \bibnamefont
  {Price}},\ }\href {\doibase 10.1103/PhysRevD.95.064028} {\bibfield  {journal}
  {\bibinfo  {journal} {Phys. Rev.}\ }\textbf {\bibinfo {volume} {D95}},\
  \bibinfo {pages} {064028} (\bibinfo {year} {2017})},\ \Eprint
  {http://arxiv.org/abs/1611.05447} {arXiv:1611.05447 [astro-ph.CO]}
  \BibitemShut {NoStop}%
\bibitem [{\citenamefont {Daverio}\ \emph {et~al.}(2016)\citenamefont
  {Daverio}, \citenamefont {Dirian},\ and\ \citenamefont
  {Mitsou}}]{Daverio:2016hqi}%
  \BibitemOpen
  \bibfield  {author} {\bibinfo {author} {\bibfnamefont {D.}~\bibnamefont
  {Daverio}}, \bibinfo {author} {\bibfnamefont {Y.}~\bibnamefont {Dirian}}, \
  and\ \bibinfo {author} {\bibfnamefont {E.}~\bibnamefont {Mitsou}},\
  }\href@noop {} {\  (\bibinfo {year} {2016})},\ \Eprint
  {http://arxiv.org/abs/1611.03437} {arXiv:1611.03437 [gr-qc]} \BibitemShut
  {NoStop}%
\bibitem [{\citenamefont {East}\ \emph {et~al.}(2018)\citenamefont {East},
  \citenamefont {Wojtak},\ and\ \citenamefont {Abel}}]{East:2017qmk}%
  \BibitemOpen
  \bibfield  {author} {\bibinfo {author} {\bibfnamefont {W.~E.}\ \bibnamefont
  {East}}, \bibinfo {author} {\bibfnamefont {R.}~\bibnamefont {Wojtak}}, \ and\
  \bibinfo {author} {\bibfnamefont {T.}~\bibnamefont {Abel}},\ }\href {\doibase
  10.1103/PhysRevD.97.043509} {\bibfield  {journal} {\bibinfo  {journal} {Phys.
  Rev.}\ }\textbf {\bibinfo {volume} {D97}},\ \bibinfo {pages} {043509}
  (\bibinfo {year} {2018})},\ \Eprint {http://arxiv.org/abs/1711.06681}
  {arXiv:1711.06681 [astro-ph.CO]} \BibitemShut {NoStop}%
\bibitem [{\citenamefont {Macpherson}\ \emph {et~al.}(2019)\citenamefont
  {Macpherson}, \citenamefont {Price},\ and\ \citenamefont
  {Lasky}}]{Macpherson:2018btl}%
  \BibitemOpen
  \bibfield  {author} {\bibinfo {author} {\bibfnamefont {H.~J.}\ \bibnamefont
  {Macpherson}}, \bibinfo {author} {\bibfnamefont {D.~J.}\ \bibnamefont
  {Price}}, \ and\ \bibinfo {author} {\bibfnamefont {P.~D.}\ \bibnamefont
  {Lasky}},\ }\href {\doibase 10.1103/PhysRevD.99.063522} {\bibfield  {journal}
  {\bibinfo  {journal} {Phys. Rev.}\ }\textbf {\bibinfo {volume} {D99}},\
  \bibinfo {pages} {063522} (\bibinfo {year} {2019})},\ \Eprint
  {http://arxiv.org/abs/1807.01711} {arXiv:1807.01711 [astro-ph.CO]}
  \BibitemShut {NoStop}%
\bibitem [{\citenamefont {Giblin}\ \emph {et~al.}(2019)\citenamefont {Giblin},
  \citenamefont {Mertens}, \citenamefont {Starkman},\ and\ \citenamefont
  {Tian}}]{Giblin:2018ndw}%
  \BibitemOpen
  \bibfield  {author} {\bibinfo {author} {\bibfnamefont {J.~T.}\ \bibnamefont
  {Giblin}}, \bibinfo {author} {\bibfnamefont {J.~B.}\ \bibnamefont {Mertens}},
  \bibinfo {author} {\bibfnamefont {G.~D.}\ \bibnamefont {Starkman}}, \ and\
  \bibinfo {author} {\bibfnamefont {C.}~\bibnamefont {Tian}},\ }\href {\doibase
  10.1103/PhysRevD.99.023527} {\bibfield  {journal} {\bibinfo  {journal} {Phys.
  Rev.}\ }\textbf {\bibinfo {volume} {D99}},\ \bibinfo {pages} {023527}
  (\bibinfo {year} {2019})},\ \Eprint {http://arxiv.org/abs/1810.05203}
  {arXiv:1810.05203 [astro-ph.CO]} \BibitemShut {NoStop}%
\bibitem [{\citenamefont {Daverio}\ \emph {et~al.}(2019)\citenamefont
  {Daverio}, \citenamefont {Dirian},\ and\ \citenamefont
  {Mitsou}}]{Daverio:2019gql}%
  \BibitemOpen
  \bibfield  {author} {\bibinfo {author} {\bibfnamefont {D.}~\bibnamefont
  {Daverio}}, \bibinfo {author} {\bibfnamefont {Y.}~\bibnamefont {Dirian}}, \
  and\ \bibinfo {author} {\bibfnamefont {E.}~\bibnamefont {Mitsou}},\
  }\href@noop {} {\  (\bibinfo {year} {2019})},\ \Eprint
  {http://arxiv.org/abs/1904.07841} {arXiv:1904.07841 [astro-ph.CO]}
  \BibitemShut {NoStop}%
\bibitem [{\citenamefont {Buchert}(2000)}]{Buchert:1999er}%
  \BibitemOpen
  \bibfield  {author} {\bibinfo {author} {\bibfnamefont {T.}~\bibnamefont
  {Buchert}},\ }\href {\doibase 10.1023/A:1001800617177} {\bibfield  {journal}
  {\bibinfo  {journal} {Gen. Rel. Grav.}\ }\textbf {\bibinfo {volume} {32}},\
  \bibinfo {pages} {105} (\bibinfo {year} {2000})},\ \Eprint
  {http://arxiv.org/abs/gr-qc/9906015} {arXiv:gr-qc/9906015 [gr-qc]}
  \BibitemShut {NoStop}%
\bibitem [{\citenamefont {Kolb}\ \emph {et~al.}(2005)\citenamefont {Kolb},
  \citenamefont {Matarrese}, \citenamefont {Notari},\ and\ \citenamefont
  {Riotto}}]{Kolb:2004am}%
  \BibitemOpen
  \bibfield  {author} {\bibinfo {author} {\bibfnamefont {E.~W.}\ \bibnamefont
  {Kolb}}, \bibinfo {author} {\bibfnamefont {S.}~\bibnamefont {Matarrese}},
  \bibinfo {author} {\bibfnamefont {A.}~\bibnamefont {Notari}}, \ and\ \bibinfo
  {author} {\bibfnamefont {A.}~\bibnamefont {Riotto}},\ }\href {\doibase
  10.1103/PhysRevD.71.023524} {\bibfield  {journal} {\bibinfo  {journal} {Phys.
  Rev.}\ }\textbf {\bibinfo {volume} {D71}},\ \bibinfo {pages} {023524}
  (\bibinfo {year} {2005})},\ \Eprint {http://arxiv.org/abs/hep-ph/0409038}
  {arXiv:hep-ph/0409038 [hep-ph]} \BibitemShut {NoStop}%
\bibitem [{\citenamefont {Rasanen}(2011)}]{Rasanen:2011ki}%
  \BibitemOpen
  \bibfield  {author} {\bibinfo {author} {\bibfnamefont {S.}~\bibnamefont
  {Rasanen}},\ }\href {\doibase 10.1088/0264-9381/28/16/164008} {\bibfield
  {journal} {\bibinfo  {journal} {Class. Quant. Grav.}\ }\textbf {\bibinfo
  {volume} {28}},\ \bibinfo {pages} {164008} (\bibinfo {year} {2011})},\
  \Eprint {http://arxiv.org/abs/1102.0408} {arXiv:1102.0408 [astro-ph.CO]}
  \BibitemShut {NoStop}%
\bibitem [{\citenamefont {Ishibashi}\ and\ \citenamefont
  {Wald}(2006)}]{Ishibashi:2005sj}%
  \BibitemOpen
  \bibfield  {author} {\bibinfo {author} {\bibfnamefont {A.}~\bibnamefont
  {Ishibashi}}\ and\ \bibinfo {author} {\bibfnamefont {R.~M.}\ \bibnamefont
  {Wald}},\ }\href {\doibase 10.1088/0264-9381/23/1/012} {\bibfield  {journal}
  {\bibinfo  {journal} {Class. Quant. Grav.}\ }\textbf {\bibinfo {volume}
  {23}},\ \bibinfo {pages} {235} (\bibinfo {year} {2006})},\ \Eprint
  {http://arxiv.org/abs/gr-qc/0509108} {arXiv:gr-qc/0509108 [gr-qc]}
  \BibitemShut {NoStop}%
\bibitem [{\citenamefont {Green}\ and\ \citenamefont
  {Wald}(2014)}]{Green:2014aga}%
  \BibitemOpen
  \bibfield  {author} {\bibinfo {author} {\bibfnamefont {S.~R.}\ \bibnamefont
  {Green}}\ and\ \bibinfo {author} {\bibfnamefont {R.~M.}\ \bibnamefont
  {Wald}},\ }\href {\doibase 10.1088/0264-9381/31/23/234003} {\bibfield
  {journal} {\bibinfo  {journal} {Class. Quant. Grav.}\ }\textbf {\bibinfo
  {volume} {31}},\ \bibinfo {pages} {234003} (\bibinfo {year} {2014})},\
  \Eprint {http://arxiv.org/abs/1407.8084} {arXiv:1407.8084 [gr-qc]}
  \BibitemShut {NoStop}%
\bibitem [{\citenamefont {{Laureijs}}\ \emph {et~al.}(2011)\citenamefont
  {{Laureijs}}, \citenamefont {{Amiaux}}, \citenamefont {{Arduini}},
  \citenamefont {{Augu{\`e}res}}, \citenamefont {{Brinchmann}}, \citenamefont
  {{Cole}}, \citenamefont {{Cropper}}, \citenamefont {{Dabin}}, \citenamefont
  {{Duvet}}, \citenamefont {{Ealet}},\ and\ \citenamefont
  {et~al.}}]{euclid2011}%
  \BibitemOpen
  \bibfield  {author} {\bibinfo {author} {\bibfnamefont {R.}~\bibnamefont
  {{Laureijs}}}, \bibinfo {author} {\bibfnamefont {J.}~\bibnamefont
  {{Amiaux}}}, \bibinfo {author} {\bibfnamefont {S.}~\bibnamefont {{Arduini}}},
  \bibinfo {author} {\bibfnamefont {J.~.}\ \bibnamefont {{Augu{\`e}res}}},
  \bibinfo {author} {\bibfnamefont {J.}~\bibnamefont {{Brinchmann}}}, \bibinfo
  {author} {\bibfnamefont {R.}~\bibnamefont {{Cole}}}, \bibinfo {author}
  {\bibfnamefont {M.}~\bibnamefont {{Cropper}}}, \bibinfo {author}
  {\bibfnamefont {C.}~\bibnamefont {{Dabin}}}, \bibinfo {author} {\bibfnamefont
  {L.}~\bibnamefont {{Duvet}}}, \bibinfo {author} {\bibfnamefont
  {A.}~\bibnamefont {{Ealet}}}, \ and\ \bibinfo {author} {\bibnamefont
  {et~al.}},\ }\href@noop {} {\bibfield  {journal} {\bibinfo  {journal} {ArXiv
  e-prints}\ } (\bibinfo {year} {2011})},\ \Eprint
  {http://arxiv.org/abs/1110.3193} {arXiv:1110.3193 [astro-ph.CO]} \BibitemShut
  {NoStop}%
\bibitem [{\citenamefont {{LSST Dark Energy Science
  Collaboration}}(2012)}]{lsst2012}%
  \BibitemOpen
  \bibfield  {author} {\bibinfo {author} {\bibnamefont {{LSST Dark Energy
  Science Collaboration}}},\ }\href@noop {} {\bibfield  {journal} {\bibinfo
  {journal} {ArXiv e-prints}\ } (\bibinfo {year} {2012})},\ \Eprint
  {http://arxiv.org/abs/1211.0310} {arXiv:1211.0310 [astro-ph.CO]} \BibitemShut
  {NoStop}%
\bibitem [{\citenamefont {Collaboration}(2017)}]{des1yr2017}%
  \BibitemOpen
  \bibfield  {author} {\bibinfo {author} {\bibfnamefont {D.}~\bibnamefont
  {Collaboration}},\ }\href@noop {} {\bibfield  {journal} {\bibinfo  {journal}
  {ArXiv e-prints}\ } (\bibinfo {year} {2017})},\ \Eprint
  {http://arxiv.org/abs/1708.01530} {arXiv:1708.01530} \BibitemShut {NoStop}%
\bibitem [{\citenamefont {Heitmann}\ \emph {et~al.}(2008)\citenamefont
  {Heitmann} \emph {et~al.}}]{Heitmann:2007hr}%
  \BibitemOpen
  \bibfield  {author} {\bibinfo {author} {\bibfnamefont {K.}~\bibnamefont
  {Heitmann}} \emph {et~al.},\ }\href {\doibase 10.1088/1749-4699/1/1/015003}
  {\bibfield  {journal} {\bibinfo  {journal} {Comput. Sci. Dis.}\ }\textbf
  {\bibinfo {volume} {1}},\ \bibinfo {pages} {015003} (\bibinfo {year}
  {2008})},\ \Eprint {http://arxiv.org/abs/0706.1270} {arXiv:0706.1270
  [astro-ph]} \BibitemShut {NoStop}%
\bibitem [{\citenamefont {Schneider}\ \emph {et~al.}(2016)\citenamefont
  {Schneider}, \citenamefont {Teyssier}, \citenamefont {Potter}, \citenamefont
  {Stadel}, \citenamefont {Onions}, \citenamefont {Reed}, \citenamefont
  {Smith}, \citenamefont {Springel}, \citenamefont {Pearce},\ and\
  \citenamefont {Scoccimarro}}]{Schneider:2015yka}%
  \BibitemOpen
  \bibfield  {author} {\bibinfo {author} {\bibfnamefont {A.}~\bibnamefont
  {Schneider}}, \bibinfo {author} {\bibfnamefont {R.}~\bibnamefont {Teyssier}},
  \bibinfo {author} {\bibfnamefont {D.}~\bibnamefont {Potter}}, \bibinfo
  {author} {\bibfnamefont {J.}~\bibnamefont {Stadel}}, \bibinfo {author}
  {\bibfnamefont {J.}~\bibnamefont {Onions}}, \bibinfo {author} {\bibfnamefont
  {D.~S.}\ \bibnamefont {Reed}}, \bibinfo {author} {\bibfnamefont {R.~E.}\
  \bibnamefont {Smith}}, \bibinfo {author} {\bibfnamefont {V.}~\bibnamefont
  {Springel}}, \bibinfo {author} {\bibfnamefont {F.~R.}\ \bibnamefont
  {Pearce}}, \ and\ \bibinfo {author} {\bibfnamefont {R.}~\bibnamefont
  {Scoccimarro}},\ }\href {\doibase 10.1088/1475-7516/2016/04/047} {\bibfield
  {journal} {\bibinfo  {journal} {JCAP}\ }\textbf {\bibinfo {volume} {1604}},\
  \bibinfo {pages} {047} (\bibinfo {year} {2016})},\ \Eprint
  {http://arxiv.org/abs/1503.05920} {arXiv:1503.05920 [astro-ph.CO]}
  \BibitemShut {NoStop}%
\bibitem [{\citenamefont {Adamek}\ \emph {et~al.}(2016)\citenamefont {Adamek},
  \citenamefont {Daverio}, \citenamefont {Durrer},\ and\ \citenamefont
  {Kunz}}]{Adamek:2015eda}%
  \BibitemOpen
  \bibfield  {author} {\bibinfo {author} {\bibfnamefont {J.}~\bibnamefont
  {Adamek}}, \bibinfo {author} {\bibfnamefont {D.}~\bibnamefont {Daverio}},
  \bibinfo {author} {\bibfnamefont {R.}~\bibnamefont {Durrer}}, \ and\ \bibinfo
  {author} {\bibfnamefont {M.}~\bibnamefont {Kunz}},\ }\href {\doibase
  10.1038/nphys3673} {\bibfield  {journal} {\bibinfo  {journal} {Nature Phys.}\
  }\textbf {\bibinfo {volume} {12}},\ \bibinfo {pages} {346} (\bibinfo {year}
  {2016})},\ \Eprint {http://arxiv.org/abs/1509.01699} {arXiv:1509.01699
  [astro-ph.CO]} \BibitemShut {NoStop}%
\bibitem [{\citenamefont {Barrera-Hinojosa}\ and\ \citenamefont
  {Li}(2019)}]{Barrera-Hinojosa:2019mzo}%
  \BibitemOpen
  \bibfield  {author} {\bibinfo {author} {\bibfnamefont {C.}~\bibnamefont
  {Barrera-Hinojosa}}\ and\ \bibinfo {author} {\bibfnamefont {B.}~\bibnamefont
  {Li}},\ }\href@noop {} {\  (\bibinfo {year} {2019})},\ \Eprint
  {http://arxiv.org/abs/1905.08890} {arXiv:1905.08890 [astro-ph.CO]}
  \BibitemShut {NoStop}%
\bibitem [{\citenamefont {Pretorius}\ and\ \citenamefont
  {East}(2018)}]{Pretorius:2018lfb}%
  \BibitemOpen
  \bibfield  {author} {\bibinfo {author} {\bibfnamefont {F.}~\bibnamefont
  {Pretorius}}\ and\ \bibinfo {author} {\bibfnamefont {W.~E.}\ \bibnamefont
  {East}},\ }\href {\doibase 10.1103/PhysRevD.98.084053} {\bibfield  {journal}
  {\bibinfo  {journal} {Phys. Rev.}\ }\textbf {\bibinfo {volume} {D98}},\
  \bibinfo {pages} {084053} (\bibinfo {year} {2018})},\ \Eprint
  {http://arxiv.org/abs/1807.11562} {arXiv:1807.11562 [gr-qc]} \BibitemShut
  {NoStop}%
\bibitem [{\citenamefont {Chisari}\ and\ \citenamefont
  {Zaldarriaga}(2011)}]{Chisari:2011iq}%
  \BibitemOpen
  \bibfield  {author} {\bibinfo {author} {\bibfnamefont {N.~E.}\ \bibnamefont
  {Chisari}}\ and\ \bibinfo {author} {\bibfnamefont {M.}~\bibnamefont
  {Zaldarriaga}},\ }\href {\doibase 10.1103/PhysRevD.84.089901,
  10.1103/PhysRevD.83.123505} {\bibfield  {journal} {\bibinfo  {journal} {Phys.
  Rev.}\ }\textbf {\bibinfo {volume} {D83}},\ \bibinfo {pages} {123505}
  (\bibinfo {year} {2011})},\ \bibinfo {note} {[Erratum: Phys.
  Rev.D84,089901(2011)]},\ \Eprint {http://arxiv.org/abs/1101.3555}
  {arXiv:1101.3555 [astro-ph.CO]} \BibitemShut {NoStop}%
\bibitem [{\citenamefont {Green}\ and\ \citenamefont
  {Wald}(2012)}]{Green:2011wc}%
  \BibitemOpen
  \bibfield  {author} {\bibinfo {author} {\bibfnamefont {S.~R.}\ \bibnamefont
  {Green}}\ and\ \bibinfo {author} {\bibfnamefont {R.~M.}\ \bibnamefont
  {Wald}},\ }\href {\doibase 10.1103/PhysRevD.85.063512} {\bibfield  {journal}
  {\bibinfo  {journal} {Phys. Rev.}\ }\textbf {\bibinfo {volume} {D85}},\
  \bibinfo {pages} {063512} (\bibinfo {year} {2012})},\ \Eprint
  {http://arxiv.org/abs/1111.2997} {arXiv:1111.2997 [gr-qc]} \BibitemShut
  {NoStop}%
\bibitem [{\citenamefont {{Zel'dovich}}(1970)}]{Zeldovich1970}%
  \BibitemOpen
  \bibfield  {author} {\bibinfo {author} {\bibfnamefont {Y.~B.}\ \bibnamefont
  {{Zel'dovich}}},\ }\href@noop {} {\bibfield  {journal} {\bibinfo  {journal}
  {Astron. Astrophys.}\ }\textbf {\bibinfo {volume} {5}},\ \bibinfo {pages}
  {84} (\bibinfo {year} {1970})}\BibitemShut {NoStop}%
\bibitem [{\citenamefont {East}\ \emph {et~al.}(2012)\citenamefont {East},
  \citenamefont {Ramazanoglu},\ and\ \citenamefont
  {Pretorius}}]{idsolve_paper}%
  \BibitemOpen
  \bibfield  {author} {\bibinfo {author} {\bibfnamefont {W.~E.}\ \bibnamefont
  {East}}, \bibinfo {author} {\bibfnamefont {F.~M.}\ \bibnamefont
  {Ramazanoglu}}, \ and\ \bibinfo {author} {\bibfnamefont {F.}~\bibnamefont
  {Pretorius}},\ }\href {\doibase 10.1103/PhysRevD.86.104053} {\bibfield
  {journal} {\bibinfo  {journal} {Phys. Rev.}\ }\textbf {\bibinfo {volume}
  {D86}},\ \bibinfo {pages} {104053} (\bibinfo {year} {2012})},\ \Eprint
  {http://arxiv.org/abs/1208.3473} {arXiv:1208.3473 [gr-qc]} \BibitemShut
  {NoStop}%
\bibitem [{\citenamefont {{Springel}}(2005)}]{Springel2005}%
  \BibitemOpen
  \bibfield  {author} {\bibinfo {author} {\bibfnamefont {V.}~\bibnamefont
  {{Springel}}},\ }\href {\doibase 10.1111/j.1365-2966.2005.09655.x} {\bibfield
   {journal} {\bibinfo  {journal} {Mon. Not. R. Astron. Soc.}\ }\textbf
  {\bibinfo {volume} {364}},\ \bibinfo {pages} {1105} (\bibinfo {year}
  {2005})},\ \Eprint {http://arxiv.org/abs/astro-ph/0505010} {astro-ph/0505010}
  \BibitemShut {NoStop}%
\bibitem [{\citenamefont {{Xu}}(1995)}]{Xu1995}%
  \BibitemOpen
  \bibfield  {author} {\bibinfo {author} {\bibfnamefont {G.}~\bibnamefont
  {{Xu}}},\ }\href {\doibase 10.1086/192166} {\bibfield  {journal} {\bibinfo
  {journal} {Astrophys. J. Supp. Ser.}\ }\textbf {\bibinfo {volume} {98}},\
  \bibinfo {pages} {355} (\bibinfo {year} {1995})},\ \Eprint
  {http://arxiv.org/abs/astro-ph/9409021} {arXiv:astro-ph/9409021 [astro-ph]}
  \BibitemShut {NoStop}%
\bibitem [{\citenamefont {{Heitmann}}\ \emph {et~al.}(2008)\citenamefont
  {{Heitmann}}, \citenamefont {{Luki{\'c}}}, \citenamefont {{Fasel}},
  \citenamefont {{Habib}}, \citenamefont {{Warren}}, \citenamefont {{White}},
  \citenamefont {{Ahrens}}, \citenamefont {{Ankeny}}, \citenamefont
  {{Armstrong}},\ and\ \citenamefont {{O'Shea}}}]{Heit2008}%
  \BibitemOpen
  \bibfield  {author} {\bibinfo {author} {\bibfnamefont {K.}~\bibnamefont
  {{Heitmann}}}, \bibinfo {author} {\bibfnamefont {Z.}~\bibnamefont
  {{Luki{\'c}}}}, \bibinfo {author} {\bibfnamefont {P.}~\bibnamefont
  {{Fasel}}}, \bibinfo {author} {\bibfnamefont {S.}~\bibnamefont {{Habib}}},
  \bibinfo {author} {\bibfnamefont {M.~S.}\ \bibnamefont {{Warren}}}, \bibinfo
  {author} {\bibfnamefont {M.}~\bibnamefont {{White}}}, \bibinfo {author}
  {\bibfnamefont {J.}~\bibnamefont {{Ahrens}}}, \bibinfo {author}
  {\bibfnamefont {L.}~\bibnamefont {{Ankeny}}}, \bibinfo {author}
  {\bibfnamefont {R.}~\bibnamefont {{Armstrong}}}, \ and\ \bibinfo {author}
  {\bibfnamefont {B.}~\bibnamefont {{O'Shea}}},\ }\href {\doibase
  10.1088/1749-4699/1/1/015003} {\bibfield  {journal} {\bibinfo  {journal}
  {Computational Science and Discovery}\ }\textbf {\bibinfo {volume} {1}},\
  \bibinfo {eid} {015003} (\bibinfo {year} {2008})},\ \Eprint
  {http://arxiv.org/abs/0706.1270} {arXiv:0706.1270 [astro-ph]} \BibitemShut
  {NoStop}%
\bibitem [{\citenamefont {{Kim}}\ \emph {et~al.}(2014)\citenamefont {{Kim}},
  \citenamefont {{Abel}}, \citenamefont {{Agertz}}, \citenamefont {{Bryan}},
  \citenamefont {{Ceverino}}, \citenamefont {{Christensen}}, \citenamefont
  {{Conroy}}, \citenamefont {{Dekel}}, \citenamefont {{Gnedin}},\ and\
  \citenamefont {{Goldbaum}}}]{Kim2014}%
  \BibitemOpen
  \bibfield  {author} {\bibinfo {author} {\bibfnamefont {J.-h.}\ \bibnamefont
  {{Kim}}}, \bibinfo {author} {\bibfnamefont {T.}~\bibnamefont {{Abel}}},
  \bibinfo {author} {\bibfnamefont {O.}~\bibnamefont {{Agertz}}}, \bibinfo
  {author} {\bibfnamefont {G.~L.}\ \bibnamefont {{Bryan}}}, \bibinfo {author}
  {\bibfnamefont {D.}~\bibnamefont {{Ceverino}}}, \bibinfo {author}
  {\bibfnamefont {C.}~\bibnamefont {{Christensen}}}, \bibinfo {author}
  {\bibfnamefont {C.}~\bibnamefont {{Conroy}}}, \bibinfo {author}
  {\bibfnamefont {A.}~\bibnamefont {{Dekel}}}, \bibinfo {author} {\bibfnamefont
  {N.~Y.}\ \bibnamefont {{Gnedin}}}, \ and\ \bibinfo {author} {\bibfnamefont
  {N.~J.}\ \bibnamefont {{Goldbaum}}},\ }\href {\doibase
  10.1088/0067-0049/210/1/14} {\bibfield  {journal} {\bibinfo  {journal}
  {Astrophys. J. Supp. Ser.}\ }\textbf {\bibinfo {volume} {210}},\ \bibinfo
  {eid} {14} (\bibinfo {year} {2014})},\ \Eprint
  {http://arxiv.org/abs/1308.2669} {arXiv:1308.2669 [astro-ph.GA]} \BibitemShut
  {NoStop}%
\bibitem [{\citenamefont {{Heitmann}}\ \emph {et~al.}(2005)\citenamefont
  {{Heitmann}}, \citenamefont {{Ricker}}, \citenamefont {{Warren}},\ and\
  \citenamefont {{Habib}}}]{Hei2005}%
  \BibitemOpen
  \bibfield  {author} {\bibinfo {author} {\bibfnamefont {K.}~\bibnamefont
  {{Heitmann}}}, \bibinfo {author} {\bibfnamefont {P.~M.}\ \bibnamefont
  {{Ricker}}}, \bibinfo {author} {\bibfnamefont {M.~S.}\ \bibnamefont
  {{Warren}}}, \ and\ \bibinfo {author} {\bibfnamefont {S.}~\bibnamefont
  {{Habib}}},\ }\href {\doibase 10.1086/432646} {\bibfield  {journal} {\bibinfo
   {journal} {Astrophys. J. Supp. Ser.}\ }\textbf {\bibinfo {volume} {160}},\
  \bibinfo {pages} {28} (\bibinfo {year} {2005})},\ \Eprint
  {http://arxiv.org/abs/astro-ph/0411795} {arXiv:astro-ph/0411795 [astro-ph]}
  \BibitemShut {NoStop}%
\bibitem [{\citenamefont {{Shandarin}}\ \emph {et~al.}(2012)\citenamefont
  {{Shandarin}}, \citenamefont {{Habib}},\ and\ \citenamefont
  {{Heitmann}}}]{Shandarin2012}%
  \BibitemOpen
  \bibfield  {author} {\bibinfo {author} {\bibfnamefont {S.}~\bibnamefont
  {{Shandarin}}}, \bibinfo {author} {\bibfnamefont {S.}~\bibnamefont
  {{Habib}}}, \ and\ \bibinfo {author} {\bibfnamefont {K.}~\bibnamefont
  {{Heitmann}}},\ }\href {\doibase 10.1103/PhysRevD.85.083005} {\bibfield
  {journal} {\bibinfo  {journal} {Phys. Rev. D}\ }\textbf {\bibinfo {volume}
  {85}},\ \bibinfo {eid} {083005} (\bibinfo {year} {2012})},\ \Eprint
  {http://arxiv.org/abs/1111.2366} {arXiv:1111.2366} \BibitemShut {NoStop}%
\bibitem [{\citenamefont {{Abel}}\ \emph {et~al.}(2012)\citenamefont {{Abel}},
  \citenamefont {{Hahn}},\ and\ \citenamefont {{Kaehler}}}]{Abel2012}%
  \BibitemOpen
  \bibfield  {author} {\bibinfo {author} {\bibfnamefont {T.}~\bibnamefont
  {{Abel}}}, \bibinfo {author} {\bibfnamefont {O.}~\bibnamefont {{Hahn}}}, \
  and\ \bibinfo {author} {\bibfnamefont {R.}~\bibnamefont {{Kaehler}}},\ }\href
  {\doibase 10.1111/j.1365-2966.2012.21754.x} {\bibfield  {journal} {\bibinfo
  {journal} {Mon. Not. R. Astron. Soc.}\ }\textbf {\bibinfo {volume} {427}},\
  \bibinfo {pages} {61} (\bibinfo {year} {2012})},\ \Eprint
  {http://arxiv.org/abs/1111.3944} {arXiv:1111.3944} \BibitemShut {NoStop}%
\bibitem [{\citenamefont {{Hahn}}\ and\ \citenamefont
  {{Angulo}}(2016)}]{Hahn2016}%
  \BibitemOpen
  \bibfield  {author} {\bibinfo {author} {\bibfnamefont {O.}~\bibnamefont
  {{Hahn}}}\ and\ \bibinfo {author} {\bibfnamefont {R.~E.}\ \bibnamefont
  {{Angulo}}},\ }\href {\doibase 10.1093/mnras/stv2304} {\bibfield  {journal}
  {\bibinfo  {journal} {Mon. Not. R. Astron. Soc.}\ }\textbf {\bibinfo {volume}
  {455}},\ \bibinfo {pages} {1115} (\bibinfo {year} {2016})},\ \Eprint
  {http://arxiv.org/abs/1501.01959} {arXiv:1501.01959 [astro-ph.CO]}
  \BibitemShut {NoStop}%
\bibitem [{\citenamefont {East}(2019)}]{East:2019bwu}%
  \BibitemOpen
  \bibfield  {author} {\bibinfo {author} {\bibfnamefont {W.~E.}\ \bibnamefont
  {East}},\ }\href {\doibase 10.1103/PhysRevLett.122.231103} {\bibfield
  {journal} {\bibinfo  {journal} {Phys. Rev. Lett.}\ }\textbf {\bibinfo
  {volume} {122}},\ \bibinfo {pages} {231103} (\bibinfo {year} {2019})},\
  \Eprint {http://arxiv.org/abs/1901.04498} {arXiv:1901.04498 [gr-qc]}
  \BibitemShut {NoStop}%
\bibitem [{\citenamefont {Etherington}(2007)}]{Etherington2007}%
  \BibitemOpen
  \bibfield  {author} {\bibinfo {author} {\bibfnamefont {I.~M.~H.}\
  \bibnamefont {Etherington}},\ }\href {\doibase 10.1007/s10714-007-0447-x}
  {\bibfield  {journal} {\bibinfo  {journal} {General Relativity and
  Gravitation}\ }\textbf {\bibinfo {volume} {39}},\ \bibinfo {pages} {1055}
  (\bibinfo {year} {2007})}\BibitemShut {NoStop}%
\bibitem [{\citenamefont {{Riess}}\ \emph {et~al.}(2019)\citenamefont
  {{Riess}}, \citenamefont {{Casertano}}, \citenamefont {{Yuan}}, \citenamefont
  {{Macri}},\ and\ \citenamefont {{Scolnic}}}]{Rie2019}%
  \BibitemOpen
  \bibfield  {author} {\bibinfo {author} {\bibfnamefont {A.~G.}\ \bibnamefont
  {{Riess}}}, \bibinfo {author} {\bibfnamefont {S.}~\bibnamefont
  {{Casertano}}}, \bibinfo {author} {\bibfnamefont {W.}~\bibnamefont {{Yuan}}},
  \bibinfo {author} {\bibfnamefont {L.~M.}\ \bibnamefont {{Macri}}}, \ and\
  \bibinfo {author} {\bibfnamefont {D.}~\bibnamefont {{Scolnic}}},\ }\href
  {\doibase 10.3847/1538-4357/ab1422} {\bibfield  {journal} {\bibinfo
  {journal} {Astrophys. J.}\ }\textbf {\bibinfo {volume} {876}},\ \bibinfo
  {eid} {85} (\bibinfo {year} {2019})},\ \Eprint
  {http://arxiv.org/abs/1903.07603} {arXiv:1903.07603 [astro-ph.CO]}
  \BibitemShut {NoStop}%
\bibitem [{\citenamefont {{Wojtak}}\ \emph {et~al.}(2014)\citenamefont
  {{Wojtak}}, \citenamefont {{Knebe}}, \citenamefont {{Watson}}, \citenamefont
  {{Iliev}}, \citenamefont {{He{\ss}}}, \citenamefont {{Rapetti}},
  \citenamefont {{Yepes}},\ and\ \citenamefont {{Gottl{\"o}ber}}}]{Wojtak2014}%
  \BibitemOpen
  \bibfield  {author} {\bibinfo {author} {\bibfnamefont {R.}~\bibnamefont
  {{Wojtak}}}, \bibinfo {author} {\bibfnamefont {A.}~\bibnamefont {{Knebe}}},
  \bibinfo {author} {\bibfnamefont {W.~A.}\ \bibnamefont {{Watson}}}, \bibinfo
  {author} {\bibfnamefont {I.~T.}\ \bibnamefont {{Iliev}}}, \bibinfo {author}
  {\bibfnamefont {S.}~\bibnamefont {{He{\ss}}}}, \bibinfo {author}
  {\bibfnamefont {D.}~\bibnamefont {{Rapetti}}}, \bibinfo {author}
  {\bibfnamefont {G.}~\bibnamefont {{Yepes}}}, \ and\ \bibinfo {author}
  {\bibfnamefont {S.}~\bibnamefont {{Gottl{\"o}ber}}},\ }\href {\doibase
  10.1093/mnras/stt2321} {\bibfield  {journal} {\bibinfo  {journal} {Mon. Not.
  R. Astron. Soc.}\ }\textbf {\bibinfo {volume} {438}},\ \bibinfo {pages}
  {1805} (\bibinfo {year} {2014})},\ \Eprint {http://arxiv.org/abs/1312.0276}
  {arXiv:1312.0276 [astro-ph.CO]} \BibitemShut {NoStop}%
\bibitem [{\citenamefont {{Wu}}\ and\ \citenamefont
  {{Huterer}}(2017)}]{Wu2017}%
  \BibitemOpen
  \bibfield  {author} {\bibinfo {author} {\bibfnamefont {H.-Y.}\ \bibnamefont
  {{Wu}}}\ and\ \bibinfo {author} {\bibfnamefont {D.}~\bibnamefont
  {{Huterer}}},\ }\href {\doibase 10.1093/mnras/stx1967} {\bibfield  {journal}
  {\bibinfo  {journal} {Mon. Not. R. Astron. Soc.}\ }\textbf {\bibinfo {volume}
  {471}},\ \bibinfo {pages} {4946} (\bibinfo {year} {2017})},\ \Eprint
  {http://arxiv.org/abs/1706.09723} {arXiv:1706.09723 [astro-ph.CO]}
  \BibitemShut {NoStop}%
\bibitem [{\citenamefont {Odderskov}\ \emph {et~al.}(2014)\citenamefont
  {Odderskov}, \citenamefont {Hannestad},\ and\ \citenamefont
  {Haugb{\o}lle}}]{Odde2014}%
  \BibitemOpen
  \bibfield  {author} {\bibinfo {author} {\bibfnamefont {I.}~\bibnamefont
  {Odderskov}}, \bibinfo {author} {\bibfnamefont {S.}~\bibnamefont
  {Hannestad}}, \ and\ \bibinfo {author} {\bibfnamefont {T.}~\bibnamefont
  {Haugb{\o}lle}},\ }\href {\doibase 10.1088/1475-7516/2014/10/028} {\bibfield
  {journal} {\bibinfo  {journal} {JCAP}\ }\textbf {\bibinfo {volume} {1410}},\
  \bibinfo {pages} {028} (\bibinfo {year} {2014})},\ \Eprint
  {http://arxiv.org/abs/1407.7364} {arXiv:1407.7364 [astro-ph.CO]} \BibitemShut
  {NoStop}%
\bibitem [{\citenamefont {{Macpherson}}\ \emph {et~al.}(2018)\citenamefont
  {{Macpherson}}, \citenamefont {{Lasky}},\ and\ \citenamefont
  {{Price}}}]{Macpherson2018}%
  \BibitemOpen
  \bibfield  {author} {\bibinfo {author} {\bibfnamefont {H.~J.}\ \bibnamefont
  {{Macpherson}}}, \bibinfo {author} {\bibfnamefont {P.~D.}\ \bibnamefont
  {{Lasky}}}, \ and\ \bibinfo {author} {\bibfnamefont {D.~J.}\ \bibnamefont
  {{Price}}},\ }\href {\doibase 10.3847/2041-8213/aadf8c} {\bibfield  {journal}
  {\bibinfo  {journal} {Astrophys. J. Lett.}\ }\textbf {\bibinfo {volume}
  {865}},\ \bibinfo {eid} {L4} (\bibinfo {year} {2018})},\ \Eprint
  {http://arxiv.org/abs/1807.01714} {arXiv:1807.01714 [astro-ph.CO]}
  \BibitemShut {NoStop}%
\bibitem [{\citenamefont {Carr}\ \emph {et~al.}(2017)\citenamefont {Carr},
  \citenamefont {Tenkanen},\ and\ \citenamefont {Vaskonen}}]{Carr:2017edp}%
  \BibitemOpen
  \bibfield  {author} {\bibinfo {author} {\bibfnamefont {B.}~\bibnamefont
  {Carr}}, \bibinfo {author} {\bibfnamefont {T.}~\bibnamefont {Tenkanen}}, \
  and\ \bibinfo {author} {\bibfnamefont {V.}~\bibnamefont {Vaskonen}},\ }\href
  {\doibase 10.1103/PhysRevD.96.063507} {\bibfield  {journal} {\bibinfo
  {journal} {Phys. Rev.}\ }\textbf {\bibinfo {volume} {D96}},\ \bibinfo {pages}
  {063507} (\bibinfo {year} {2017})},\ \Eprint
  {http://arxiv.org/abs/1706.03746} {arXiv:1706.03746 [astro-ph.CO]}
  \BibitemShut {NoStop}%
\bibitem [{\citenamefont {Braden}\ \emph {et~al.}(2017)\citenamefont {Braden},
  \citenamefont {Johnson}, \citenamefont {Peiris},\ and\ \citenamefont
  {Aguirre}}]{Braden:2016tjn}%
  \BibitemOpen
  \bibfield  {author} {\bibinfo {author} {\bibfnamefont {J.}~\bibnamefont
  {Braden}}, \bibinfo {author} {\bibfnamefont {M.~C.}\ \bibnamefont {Johnson}},
  \bibinfo {author} {\bibfnamefont {H.~V.}\ \bibnamefont {Peiris}}, \ and\
  \bibinfo {author} {\bibfnamefont {A.}~\bibnamefont {Aguirre}},\ }\href
  {\doibase 10.1103/PhysRevD.96.023541} {\bibfield  {journal} {\bibinfo
  {journal} {Phys. Rev.}\ }\textbf {\bibinfo {volume} {D96}},\ \bibinfo {pages}
  {023541} (\bibinfo {year} {2017})},\ \Eprint
  {http://arxiv.org/abs/1604.04001} {arXiv:1604.04001 [astro-ph.CO]}
  \BibitemShut {NoStop}%
\bibitem [{\citenamefont {Schwarz}\ \emph {et~al.}(2016)\citenamefont
  {Schwarz}, \citenamefont {Copi}, \citenamefont {Huterer},\ and\ \citenamefont
  {Starkman}}]{Schwarz2016}%
  \BibitemOpen
  \bibfield  {author} {\bibinfo {author} {\bibfnamefont {D.~J.}\ \bibnamefont
  {Schwarz}}, \bibinfo {author} {\bibfnamefont {C.~J.}\ \bibnamefont {Copi}},
  \bibinfo {author} {\bibfnamefont {D.}~\bibnamefont {Huterer}}, \ and\
  \bibinfo {author} {\bibfnamefont {G.~D.}\ \bibnamefont {Starkman}},\ }\href
  {\doibase 10.1088/0264-9381/33/18/184001} {\bibfield  {journal} {\bibinfo
  {journal} {Class. Quant. Grav.}\ }\textbf {\bibinfo {volume} {33}},\ \bibinfo
  {pages} {184001} (\bibinfo {year} {2016})},\ \Eprint
  {http://arxiv.org/abs/1510.07929} {arXiv:1510.07929 [astro-ph.CO]}
  \BibitemShut {NoStop}%
\bibitem [{\citenamefont {{LSST Science Collaboration}}\ \emph
  {et~al.}(2009)\citenamefont {{LSST Science Collaboration}} \emph
  {et~al.}}]{LSST2009}%
  \BibitemOpen
  \bibfield  {author} {\bibinfo {author} {\bibnamefont {{LSST Science
  Collaboration}}} \emph {et~al.},\ }\href@noop {} {\bibfield  {journal}
  {\bibinfo  {journal} {arXiv e-prints}\ } (\bibinfo {year} {2009})},\ \Eprint
  {http://arxiv.org/abs/0912.0201} {arXiv:0912.0201 [astro-ph.IM]} \BibitemShut
  {NoStop}%
\end{thebibliography}%

\end{document}